\documentclass[fleqn,usenatbib]{mnras}

\usepackage{newtxtext,newtxmath}

\usepackage[T1]{fontenc} 
\usepackage{ulem}

\DeclareRobustCommand{\VAN}[3]{#2}
\let\VANthebibliography\thebibliography
\def\thebibliography{\DeclareRobustCommand{\VAN}[3]{##3}\VANthebibliography}

\usepackage{graphicx}	
\usepackage{amsmath}	
\usepackage{movie15}
\usepackage{url}
\usepackage{xcolor}
\let\oldAA\AA
\renewcommand{\AA}{\text{\normalfont\oldAA}}

\newcommand{\ms}{$M_{\star}$}

\newcommand{\msun}{M$_{\sun}$}
\newcommand{\re}{$R_{e}$}

\newcommand{\rer}{$R_{e,r}$}

\newcommand{\rem}{$R_{\rm e,M}$}
\newcommand{\rmass}{$R_{\rm 50M}$}
\def\rlight{\mbox{$R_{\rm 50L}$}}
\newcommand{\rlightr}{$R_{50r}$}
\newcommand{\tlb}{$t_{\rm lb}$}

\def\agemw{\mbox{Age$_{\rm mw}$}}
\def\zmw{\mbox{Z$_{\rm mw}$}}

\bibpunct[; ]{(}{)}{;}{a}{}{;}

\title[Evolution of radial gradients of E galaxies]{The evolution of radial gradients of MaNGA quiescent elliptical galaxies: inside-out quenching or outer mass growth?}
\author[V. Avila-Reese et al.]{
V.~ Avila-Reese,$^{1}$\thanks{E-mail: avila@astro.unam.mx (VAR)}
H.~Ibarra-Medel,$^{2,3}$
I.~ Lacerna,$^{3,4}$
A.~ Rodr\'iguez-Puebla,$^{1}$ \and
J. A. V\'azquez-Mata,$^{5}$
S.~ F.~ S\'anchez,$^{1}$
H.~ M.~ Hern\'andez-Toledo,$^{1}$
C. Cannarozzo$^{1}$
\\
$^{1}$Instituto de Astronom\'{\i}a, Universidad Nacional Aut\'onoma de M\'exico, A.P. 70-264, 04510 M\'exico D. F., M\'exico \\
$^{2}$Escuela Superior de F\'{\i}sica y Matem\'aticas, Instituto Polit\'ecnico Nacional, U.P. Adolfo L\'opez Mateos, C.P. 07738, Ciudad de M\'exico, M\'exico \\
$^{3}$Instituto de Astronom\'ia y Ciencias Planetarias, Universidad de Atacama, Copayapu 485, Copiap\'o, Chile\\
$^{4}$Millennium Institute of Astrophysics (MAS), Nuncio Monse\~nor S\'otero Sanz 100, Providencia, Santiago, Chile \\
$^{5}$Departamento de F\'isica, Facultad de Ciencias, Universidad Nacional Aut\'onoma de M\'exico, Ciudad Universitaria, CDMX, 04510, M\'exico
}

\date{Accepted XXX. Received YYY; in original form ZZZ}

\pubyear{2015}

\begin{document}
\label{firstpage}
\pagerange{\pageref{firstpage}--\pageref{lastpage}}
\maketitle

\begin{abstract}
Using spatially-resolved fossil record analysis on a large sample of 'red and dead' elliptical galaxies (classical ellipticals, CLEs) from the MaNGA/SDSS-IV DR15 survey, we reconstruct the archaeological evolution of their radial gradients in mass-to-luminosity ratio ($M/L$), $g-r$ color, and specific star formation (SF) rate. We also calculate other metrics that quantify the inside-out SF quenching and external mass growth processes. The $M/L$ gradients, $\nabla\Upsilon_{\star}$, are approximately flat at high look-back times (\tlb), but then they become negative and steeper until an epoch, when this trend reverses. These trends are shifted to later epochs the less massive the galaxies are. Color gradients follow qualitatively similar trends. We find that these trends are mainly driven by strong inside-out quenching, without significant outer growth or structural changes overall. 
Our results suggest a scenario where the main progenitors of local CLE galaxies evolved quasi-passively after an early dissipative phase, but underwent radial photometric changes due to the inside-out quenching that led to the systematic decrease of $\nabla\Upsilon_{\star}$ and to an increase of the light-weighted radius. The late reversing of $\nabla\Upsilon_{\star}$, \tlb$\approx2-4$ Gyr, roughly coincides with the global quenching of the CLE galaxies.
We have pushed archaeological inferences to the limit, but thanks to the large number of objects and an understanding of how the caveats and assumptions affect our results, we conclude that they offer an average description of evolutionary behaviors of CLE progenitors that is valid at least qualitatively.

\end{abstract}



\section{Introduction}

Understanding the nature and evolution of Elliptical (E) galaxies is crucial to constrain models of galaxy formation and evolution: these galaxies hold fossil information related to the star formation (SF) and metal enrichment activity that occurred in the early Universe \citep[e.g.,][]{Renzini2006,Thomas+2010,Matteucci2012,McDermid+2015,Citro+2016}, as well as to the processes of subsequent quenching of SF activity \citep[][]{Daddi+2005,Citro+2016,Man-Belli2018}. Furthermore, E galaxies, especially the massive ones, are located in the highest-density and most spatially clustered regions of the Universe \citep[][]{Dressler1980,Balogh+1999}. 
The popular two-phase formation scenario for early-type galaxies posits that at early times ($z\gtrsim 2$), an \textit{in situ} compact stellar component forms rapidly through strong gas dissipation processes and intense bursts of SF in the centers of dense and highly clustered dark matter haloes, and latter on ($z< 2$), further accretion through dry mergers (adding stars ex situ) shapes their structure and kinematics \citep[][]{Bezanson+2009,Naab+2009,Oser+2010,Hilz+2013,Rodriguez-Gomez+2016}.

According to the two-phase scenario, in the first (dissipative) phase the early-type galaxies substantially and rapidly grow in stellar mass, while in the second (non-dissipative) phase, minor mergers make them grow efficiently in size, settling the accreted stellar population prevalently in the outer region of the main progenitor. The latter is partially sustained by observations of selected samples of early-type or quiescent galaxies in surveys at different redshifts
\citep[e.g.,][for a brief review and more references, see \citealp{Ibarra-Medel+2022}]{Trujillo+2006,vanDokkum+2010,Cimatti+2012}. In fact, these studies do not follow the evolution of individual galaxies, but rather infer average evolutionary trends of some properties for selected galaxy populations at different redshifts. The results depend critically on the way in which the selection is carried out, which seeks to ensure that these populations correspond to the evolutionary sequence of the same type of galaxies on average.
Given this uncertainty, as discussed in \citet[][hereinafter Paper I]{Ibarra-Medel+2022}, two scenarios are possible for the evolution of present-day early-type galaxies after the dissipative phase. One supports the inside-out mass growth 
and another suggests a quasi-passive evolution, where the apparent size growth inferred from look-back studies can be explained by different effects: 
differences in the used samples at different redshifts
and the redshift-dependent selection effect called \textit{progenitor bias} \citep[][]{Cassata+2013,Belli+2015,Carollo+2016,Fagioli+2016,Gargiulo+2017,Faisst+2017,Zanisi+2021,Ji-Giavalisco2022a}; 
systematic uncertainties and methodological assumptions on light radial profiles  \citep[e.g.,][]{vanDokkum+2008,vanderWel+2009,Ribeiro+2016,Mosleh+2017,Genel+2018,Roy+2018,Whitney+2019,Miller+2022}; and changes over redshift of the stellar mass-to-light ratio ($M/L$) and colour gradients  \citep[e.g.,][]{Suess+2019a,Suess+2019b}. 
The latter authors, using data from the CANDELS survey \citep[][]{Grogin2011ApJS, Koekemoer2011ApJS}, have shown that the observed half-light radius of quiescent galaxies becomes systematically larger than the half-mass radius with time. The ratio of these radii is commonly used as an indicator of colour or $M/L$ gradients.  
Therefore, evolution of the $M/L$ or colour gradients may mimic strong evolution in the light size--\ms\ distribution, as shown in \citet[][]{Suess+2019a,Suess+2019b}, \citet[][]{Mosleh+2020}, and \citet[][]{Miller+2022}.

A different approach to galaxy evolution is \textit{the fossil record or archaeological method} based on spectral ''inversion'' using stellar population synthesis (SPS) models \citep[see for a review,][]{Conroy2013}. The application of this method to spatially-resolved integral field spectroscopy (IFS) data allows us to reconstruct the local and global SF, stellar mass, and chemical histories of galaxies \citep[for a recent review, see][and more references therein]{Sanchez2020}. 
The advantage of this method is that we can reconstruct \textit{individually} these evolutionary features of galaxies, both globally and locally.
However, the method suffers from their own limitations. On the one hand, the spectral inversion using the SPS models becomes increasingly uncertain to infer ages for old systems \citep[$\gtrsim 8-9$ Gyr, see][]{Conroy2013} because the spectra of stellar populations older than these ages are very similar among them. { Furthermore, archaeological inferences are subject to other uncertainties and systematic degeneracies.}  On the other hand, fossil record reconstruction lacks information about where the stellar populations assembled, either in the place where they are observed or far from there, even in separate systems that finally merged. 

In Paper I, we have applied the fossil record code \verb|Pipe3D| \citep[][]{Sanchez+2016_p21} to a large sample of ``dead and red'' (classical) E galaxies from the ``Mapping Near Galaxies at APO'' (MaNGA) survey \citep[][]{Bundy+2015}. Following the {\it individual} evolution of the 2D spatial distributions of stellar light and mass, 
we have shown that the ratio of half-mass to half-light radius decreases on average with decreasing redshift,\footnote{Due to the limited field of view of MaNGA observations, these characteristic radii are slightly smaller than those obtained from photometric observations (see Fig. 1 in Paper I), but one expect that both radii are similarly affected by the aperture limit such that their ratios are not affected.} in agreement with the results of \citet[][]{Suess+2019a,Suess+2019b} and \citet[][]{Miller+2022} obtained from look-back observations of selected samples from the CANDELS survey. However, at low redshifts, in between $z\approx0.2$ and $z\approx 0.35$, where the CANDELS survey becomes highly incomplete, our results show that this ratio reaches a minimum (an upturn) after which it increases.  We have found a similar behavior for the ratio of light to mass concentration. 
Radial variations in the $M/L$ ratio ($\Upsilon_{\star}$) or colour cause that the galaxy's light profile is different from the mass one such that any inferred structural evolution based on photometric studies differs from the (stellar mass) intrinsic evolution \citep[][see also \citealp{LaBarbera-Carvalho2009,Kennedy+2015,Ciocca+2017,Marian+2018}]{Suess+2019a,Suess+2019b,Suess+2020,Miller+2022,Hasheminia+2022}. If $\Upsilon_{\star}$ has a negative gradient that steepens with time, then the half-light radius becomes systematically larger than the half-mass radius, and if it decreases, then the former radius approaches to the latter. 
Thus, the results presented in Paper I should be consequence of the $M/L$ gradient evolution of the studied classical E galaxies (CLEs), namely ``red and dead'' E galaxies. 

Here, we present fossil record inferences for the evolution of the radial distributions of both $\Upsilon_{\star}$ (in the $g,$ $r,$ and $i$ rest-frame bands) and $g-i$ colour for the sample of MaNGA CLE galaxies presented in Paper I (see also \citealp{Lacerna+2020}), and explore what drives this evolution. In Paper I we speculate that the systematic decrease in the mass-to-light radius ratio (or in the light-to-mass concentration ratio) is due to inside-out SF quenching, although inside-out growth, especially in the outer regions, could also contribute to evolution of this ratio.  Regarding the late upturn in the mass-to-light radius ratio, we interpreted this as a result of the long-term and global SF quenching that CLE galaxies suffered, so that the stellar populations get old even in the outermost regions in such a way that the light radial distribution tends to that of the stellar mass distribution.

Which of the two processes mentioned above dominates in the evolution of the ratio of half-mass to half-light radius and of the $M/L$ gradient of CLE galaxy progenitors? The inside-out SF quenching primarily involves radial photometric evolution in a quasi-passive regime while the inside-out mass growth involves an structural change in galaxies. Here, using the fossil record method, we explore the above question as a function of stellar mass, and discuss their possible implications, as well as the caveats implied in our inferences.

The content of the paper is as follows. In Section \ref{sec:data} we describe the sample selection and the tools that we use for analysis. The results on the evolution of the $M/L$ and colour gradients are presented in Section \ref{sec:results} together with their interpretations for the evolution of the progenitors of local CLE galaxies. Section \ref{sec:quenching-vs-growth} explores what drives the $M/L$ gradient evolution of our CLE galaxies, inside-out SF quenching or external mass growth. In Section \ref{sec:scatter} we explore the randomness in the evolution of the $M/L$ and specific star formation rate (sSFR) gradients of individual galaxies and how much they contribute to the total scatter around the median trajectories in mass bins presented in the previous sections. The caveats of our methodology and how they affect our results are widely discussed in Section \ref{sec:caveats}. Section \ref{sec:implications} is devoted to discuss the possible implications of our results. Finally, in Section \ref{sec:conclusions} we present a summary and the conclusions of our study.

{ In this paper, we assume a cosmological model with flat geometry, $h=0.71$, $\Omega_m=0.27$, and $\Omega_\Lambda=0.73$, as well as a universal initial mass function (IMF) described by the \citet[][]{Salpeter+55} function.}  


\section{The sample and analysis tools} 
\label{sec:data}

Our sample of  ``red and dead'' E galaxies, the CLEs, is the same that was presented in Paper I and it is based on the MaNGA data from the Sloan Digital Sky Survey (SDSS) Data Release 15 \citep[DR15;][]{DR15+2019}, which contains 4621 galaxies. 
MaNGA is an IFS survey \citep[][]{Bundy+2015} that recently completed the observation of $\sim 10,000$ galaxies in the redshift range of 0.01 $<$ $z$ $<$ 0.15 in the wavelength range 3600--10300 \AA, and a median spatial resolution of 2.54 arcsec FWHM 
\citep[1.8 kpc at the median redshift of 0.037,][]{Drory+2015,Law+2015,Law+2016DRP} given by the atmospheric and instrumental point spread function, PSF.

As mentioned in Paper I, we use the morphological classification of MaNGA DR15 galaxies from \citet[][]{Vazquez-Mata+2022}, which is part of one of the official MaNGA Value Added Catalogs,\footnote{\url{https://www.sdss.org/dr17/data_access/value-added-catalogs/?vac_id=manga-visual-morphologies-from-sdss-and-desi-images}} and it is reported in \citet{DR17+2021}. The morphological classification in \citet[][]{Vazquez-Mata+2022} is based on the visual inspection of a combination of 
(i) newly background-subtracted and gradient-removed SDSS $r$-band and $gri$ colour images and (ii) post-processed deep $r$-band images, $grz$ colour images and PSF-deconvolved residual (after best model subtraction) images from the DESI Legacy Imaging Surveys \citep[][]{Dey+2019}. The use of the deep DESI images and the visual detection of weak rings and sharp surface brightness edges, allowed the authors to separate E, S0, and S0a candidates as much as possible \citep[for details, see][]{Vazquez-Mata+2022}.

For the SDSS DR15, a total of 722 galaxies were classified as Es, after excluding those with strong merger tidal features, bright clumps or extended objects within the MaNGA FoV (137; see for details, \citealp[][]{Vazquez-Mata+2022}). Notice that some of the galaxies  classified as Es may contain a small and faint inner disk-like structure, which according to some authors may lead to classify them as \textit{Elliculars} \citep[e.g.,][]{Graham2019}. 

For the stellar mass, \ms, we use the NSA catalog \citep[][\citealp{Chabrier2003} IMF]{Blanton+2005}; these masses were calculated using the SDSS photometry, which is not limited by fixed apertures as in the case of MaNGA, and were widely used in numerous previous works. The size of our CLE galaxies are characterized by the circularized effective radius in the $r$-band calculated as \rer $= {\it a}_{e,r}\times \sqrt{(b/a})_r$, where $a_{e,r}$ is the half-light semi-major axis and $(b/a)_r$ is the minor-to-major axis ratio. 
We take these quantities from the MaNGA PyMorph DR15 photometric catalog \citep[][]{Fischer+2019}\footnote{\url{https://data.sdss.org/datamodel/files/MANGA_PHOTO/pymorph/PYMORPH_VER/manga-pymorph.html}}.
Their case of S\'ersic truncated profile has been used. We removed 39 Es in which the S\'ersic fit failed due to contamination, peculiarity, bad-image or bad model fit. 
In addition, we removed four galaxies with a stellar mass lower than the completeness limit given in \citet[][for details, see their appendix C]{Rodriguez-Puebla+2020}.

In Paper I we did not consider other 25 E galaxies with \rer{} smaller than the MaNGA $r$-band PSF radius, that is, those with \rer/$r_{\rm PSF}\le1.0$. {This radius, $r_{\rm PSF}\approx1.25''$, is half of the FWHM PSF; although the observational seeing is $\approx 1.5''$, the reconstructed PSF in combined MaNGA datacubes after dithering and fiber sampling is $\approx 2.5''$ (FWHM).
For this work, we decided to consider only galaxies with \rer/$r_{\rm PSF}\gtrsim 1.75$, which eliminates 20 galaxies more than in Paper I, mostly of low masses. This criterion is a compromise to ensure that the effects of the PSF on our results are minor (see Sect. \ref{sec:spatial-resolution}), but at the same time still have a reasonable number of objects at low masses. 
} 
Therefore, our final pruned sample of Es is of { 634} galaxies, which we further subclassify as discussed below. 


As in Paper I, and following \citet[][]{Lacerna+2020}, we characterize MaNGA galaxies by colour, SF activity, and light-weighted stellar age. Our study is focused on red, retired (quenched) E galaxies, the CLEs, that is, those Es classified as: \textit{(i)} red according to the criterion found by \citet{Lacerna+2014} in the $g-i$ colour vs. \ms\ diagram;\footnote{Colours were taken from the SDSS database with extinction corrected \textit{modelMag} magnitudes.} \textit{(ii)} quenched by using the criterion EW(H$\alpha$)$<3\AA$ \citep[]{Sanchez+2014, Cano-Diaz+2019}, where EW(H$\alpha$) is the Pipe3D integral extinction-corrected H$\alpha$ equivalent width; and \textit{(iii)} old, with luminosity-weighted ages $>$ 4 Gyr. As in Paper I, galaxies with active galactic nuclei were excluded. The above criteria are met by { 523} 
of the pruned E galaxy sample ($82.5$\% of the total). This is the sample of CLE galaxies that we study in this paper.

\begin{figure}
	\includegraphics[width=\columnwidth]{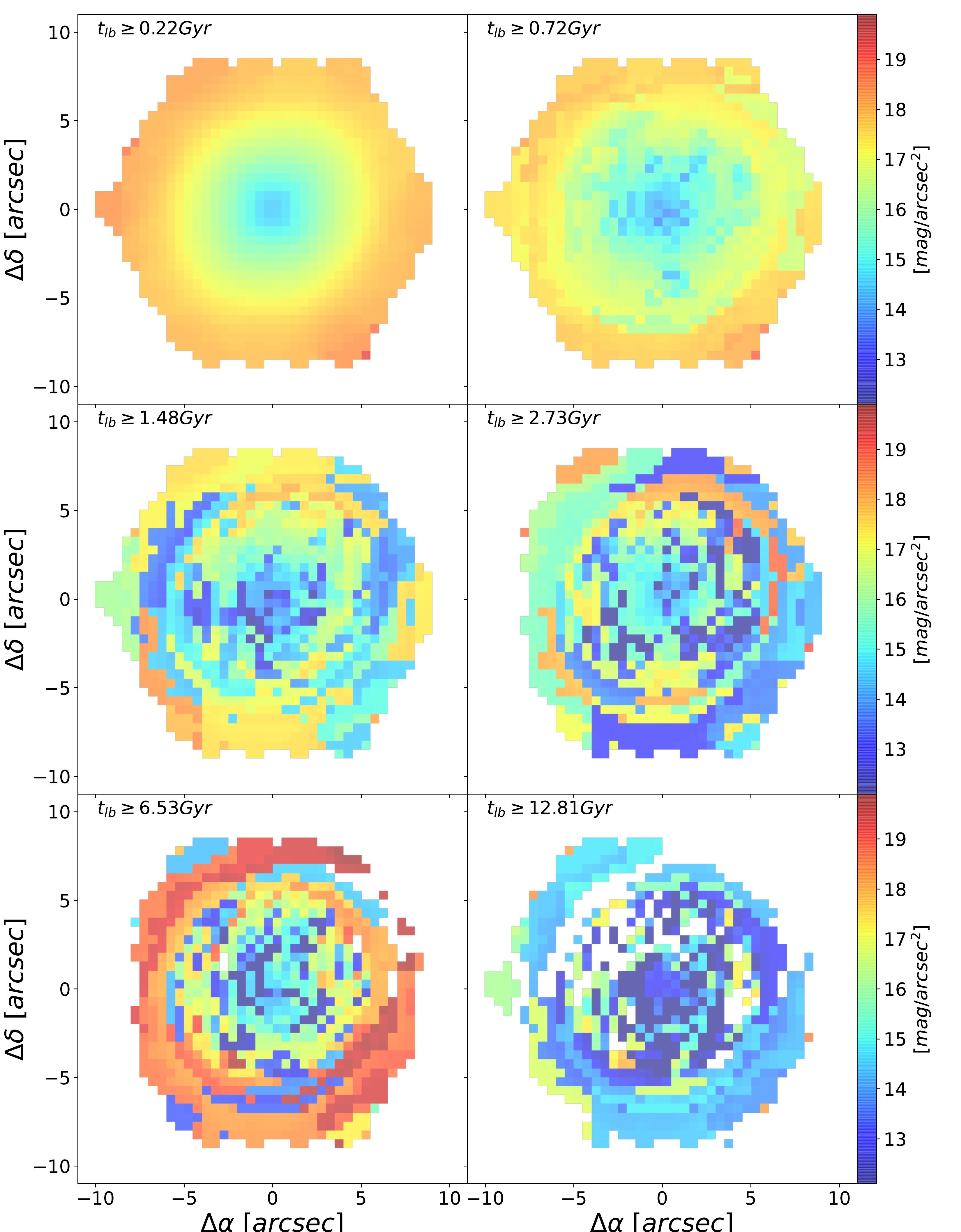}
    \caption{Rest-frame $g$-band surface brightness as would be observed { (integrating all the stellar populations up to the given look-back time)} at six selected epochs for the MaNGA galaxy with ID 1-384097 (see text). This is the type of 2D maps used here to calculate the radial gradients at different epochs of  $\Upsilon_{\star}$, color, sSFR, and surface mass density. 
    }
    \label{fig:surf-mass-evol}
\end{figure}

\begin{figure}
	\includegraphics[width=\columnwidth]{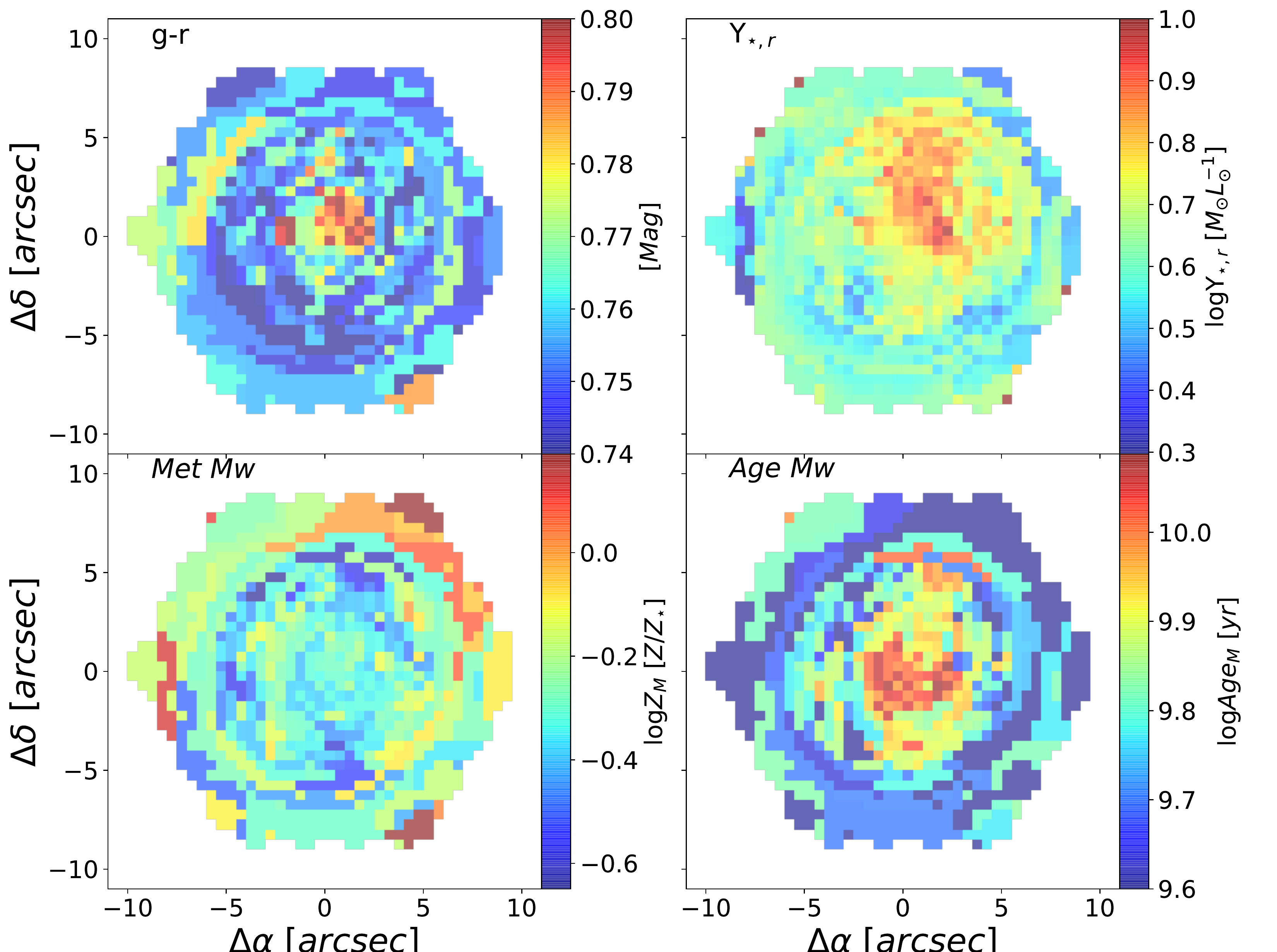}
    \caption{Color, $\Upsilon_{\star, r}$, mass-weighted metallicity, and mass-weighted age 2D maps for the MaNGA galaxy with ID 1-384097 at the observation redshift, $z_{\rm obs}=0.016$.
    }
    \label{fig:maps}
\end{figure}

We apply to the MaNGA CLE galaxies the same spatially resolved fossil record analysis as in Paper I. This analysis was performed using the \verb|Pipe3D| pipeline 
\citep{Sanchez+2016_p21,Sanchez+2018_AGN}. \verb|Pipe3D| is based on the code \verb|FIT3D| \citep{Sanchez+2016_p21,Sanchez+2016_p171} to model and subtract the stellar spectrum and fit the emission lines. \verb|FIT3D| adopts the gsd156 single stellar population (SSP) library from \citet[][]{CidFernandez+2013}. This library results from the combination of synthetic SSP spectra generated with the \verb|MILES| empirical stellar library \citep[][]{Vazdekis:2010aa} and the \verb|GRANADA| theoretical stellar library \citep{Gonzalez-Delgado+2005,Gonzalez-Delgado+2010}. The SSP templates use a \citet{Salpeter+55} IMF and cover 39 stellar ages, from 1 Myr to 14.1 Gyr (in practice, we use results until 13 Gyr), and four metallicities, $Z_\star=0.002$, 0.008, 0.02, and 0.03. \verb|FIT3D| considers the effects of dust extinction in the stellar population synthesis analysis using a \citet{Cardelli+1989} extinction law.

\verb|Pipe3D| provides non-parametric SSP decomposition of the modeled stellar spectra of each spatially resolved region of the MaNGA galaxies from which we calculate the spatially-resolved SSP age distributions (a proxy of the SF histories). From this analysis, we obtain 2D maps of stellar mass, light, $M/L$ ratio, rest-frame colours, metallicity, etc. \textit{integrated up to a given SSP age}. From that we reconstruct the spectrum of each spaxel at rest as a function of the SSP age; for details, see \S\S\ 2.1.1 in Paper I. As in that paper, here we use the SDSS $g,$ $r,$ and $i$ bands, with their corresponding response functions. 
As an example, Fig. \ref{fig:surf-mass-evol}  shows the rest-frame $g$-band surface brightness maps at different ages for one of our CLE galaxies. {  Note that these maps are not for the stellar populations corresponding to a time slice (differential) but rather integrate all stellar populations from the oldest to those corresponding to the given age.} In the past, the progenitor of this CLE galaxy has been of higher surface brightness. Figure \ref{fig:maps} shows for the same galaxy, the 2D maps of $g-r$ colour, $\Upsilon_{\star, r}$, light-weighted stellar metallicty and age at the observation time. The azimutally-averaged radial gradients are roughly flat or slightly negative.

To study the change with cosmic time of the stellar mass and light radial distribution (and its ratio, $\Upsilon_\star$) we use the 2D maps of stellar light and mass, gradually excluding stellar populations from younger to older.  We pass from the SSP ages to a common cosmic look-back time, \tlb, such that
\begin{equation}
    $\tlb$ = t_{\rm lb,obs} + Age_{\rm SSP},
\end{equation}
 where $t_{\rm lb,obs}$ is the look-back time corresponding to the observational redshift of a given galaxy 
 and $Age_{\rm SSP}$ is the respective SSP age.
To calculate the radial distributions, we use elliptical annuli along the semi-major axis. In this sense, the half-luminosity and half-mass semi-major axis at each \tlb\ are calculated as those where half of the total luminosity or mass within the FoV are attained at that \tlb.
The above semi-major axis are circularized and corrected roughly by the PSF to calculate the half-light and half-mass radii, \rlight\ and \rmass,  at each \tlb, see Paper I. For example, in the $r$ band, 
\begin{equation}
R_{50,r}($\tlb$) =  a_{50,r}($\tlb$)\times\sqrt{(b/a})_r, 
\end{equation}
where 
\begin{equation}
 a_{50,r}$(\tlb)=$\sqrt{a_{50,r,{\rm obs}}^2{(t_{\rm lb})} - r_{\rm PSF}^2}.
\end{equation}
Note that we use the subscript $e$ for the effective radii from the SDSS photometry, while the subscript 50 for the half-light radii calculated from the fossil record 2D maps {\it within the FoVs of the MaNGA observations}. 
Due to the limitation of the FoVs of MaNGA galaxies,\footnote{The MaNGA FoVs cover up to $\sim1.5$\rer\ of the galaxy in the Primary+ sample ($60.1\%$ of the CLE sample) and up to $\sim2.5$\rer\ in the Secondary sample ($36.3\%$ per cent of the CLE sample). The remaining 3.6\% corresponds to ancillary and commissioning CLE galaxies.}
the half-light (or half-mass) radii calculated from the MaNGA data cubes are smaller than those measured from the SDSS photometry (see lower panel of Fig. 1 in Paper I).

\section{Evolution of the mass-to-light and colour gradients}
\label{sec:results}

In this Section we present the variation with look-back time (or $z$) of $M/L$ and $g-i$ colour gradients for our MaNGA CLE galaxies (\S\S\ \ref{sec:m-l_gradients}),
as well as the change of the inner and outer values of $\Upsilon_\star$ with the aim to explore what drives the evolution of its radial gradient (\S\S\ \ref{sec:inside-out}). 
It is important to note that the inferences presented here refer to the \textit{progenitor} evolution of the observed CLE MaNGA galaxies, and that the progenitors may not have been a red, quiescent E galaxy in the not too distant past \citep[e.g.,][]{Lacerna+2020}, although for the most massive galaxies, this effect should be small since $z\sim 0.7$, or as look-back studies have shown, since $z\sim 1$ \citep[e.g.,][]{Lopez-SanJuan+2012}. Furthermore, a fraction of the observed stellar populations in CLE galaxies may have not been formed in the progenitor, but may have come from accreted galaxies.

\begin{figure*}
    \includegraphics[width=2\columnwidth]{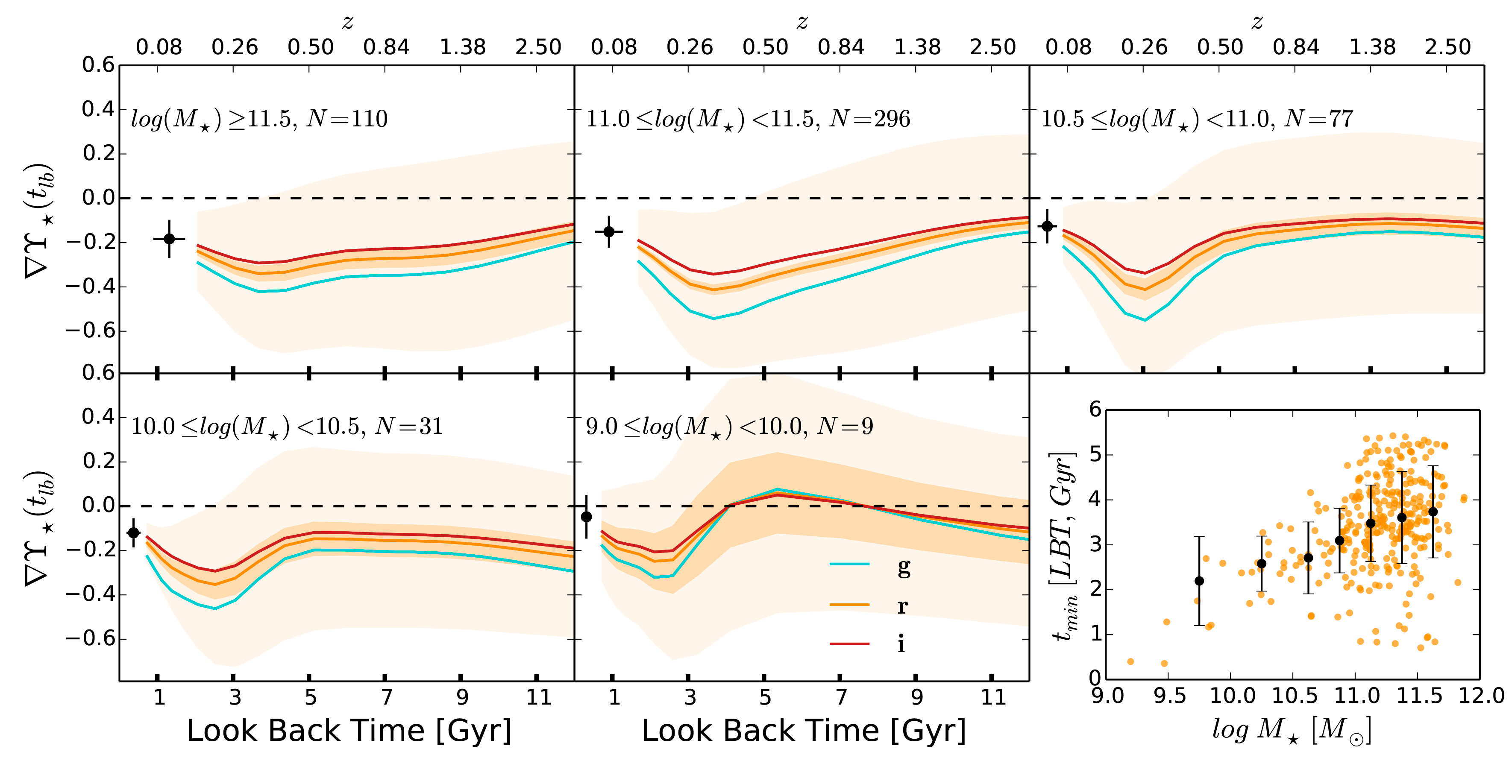}
    \caption{Evolution of the $M/L$  gradient of the MaNGA CLE galaxies. Each panel shows the running medians of the gradients for the $g,$ $r,$ and $i$ bands in the \ms($z_{\rm obs})$\ bins shown inside the panels (in units of \msun). The light shaded regions correspond to the associated 25th-75th percentiles for the $r$ band, while the dark shaded regions are estimates of the standard error of the median. By comparing it with Fig. 2 from Paper I, we see that the evolution of the \rmass-to-\rlight\ ratio follows closely the evolution of the $\Upsilon_\star$ gradient. 
    Black dots are the { median} $t_{\rm lb,obs}$ and $\nabla\Upsilon_{\star,r}$ of the observed galaxies within the given mass bin, with error bars showing the respective { 25th-75th percentiles. The lower right inset shows the look-back time at which each (smoothed) $\nabla\Upsilon_{\star,r}$ trajectory reaches its minimum, $t_{\rm min}$,  vs.  \ms($z_{\rm obs})$. The dots with error bars are the means and standard deviations in different mass bins.}
    }
    \label{fig:mass-light-grad-evol}
\end{figure*}

\subsection{The mass-to-light and colour gradient evolution}
\label{sec:m-l_gradients}

For each CLE galaxy, we calculate its $M/L$ gradient, $\nabla\Upsilon_\star$, { as it would be observed } at a given epoch as follows. For a given look-back time, we measure the integrated { up to that time} mass-to-light ratios (in the $g,$ $r$, and $i$ bands) within an inner and an outer radial regions of corresponding characteristic radii $R^{\rm in}$ and $R^{\rm out}$, $\Upsilon_\star^{\rm in}$ and $\Upsilon_\star^{\rm out}$, respectively. The gradient is then calculated as:
\begin{equation}
    \nabla\Upsilon_\star=\log(\Upsilon_\star^{\rm in}/\Upsilon_\star^{\rm out})/\log(R^{\rm in}/R^{\rm out}).
\end{equation}
The inner region is defined as an ellipse with semi-major axis corresponding to 0.5 $\mathcal{R}_{50}$(\tlb), while the outer region is an annulus within the semi-major axes corresponding to 1.2 and 1.5 $\mathcal{R}_{50}$(\tlb), where $\mathcal{R}_{50}$(\tlb)= \rlightr(\tlb)$\times$\rer/\rlightr($t_{\rm lb,obs}$).
Note that most of galaxies come from the Primary+ sample, which is limited to a FoV of $1.5$ \rer. The widths of the inner and outer regions were selected to minimize the cases where these widths are larger than the MaNGA PSF size. For the small fraction of cases where this still happens, the widths are extended to the PSF size. For the values of the mean radii characterizing the inner and outer regions, we use $R^{\rm in}=0.3\mathcal{R}_{50}$ and $R^{\rm out}=1.3\mathcal{R}_{50}$, respectively, since these are typically the mean radii of the spaxels within the defined regions.
We have experimented with other ways of measuring gradients: performing a log-linear fit to all spaxel values within 1.5 $\mathcal{R}_{50}$(\tlb) or calculating integrated values in many annuli up to 1.5 $\mathcal{R}_{50}$(\tlb) and performing a log-linear fit to these values. We have found that our main results remain nearly the same with these different ways of calculating radial gradients but we prefer the selected one because it provides more robust determinations of the gradients (less dispersion along time for a given galaxy and less scatter for the population at a given \tlb\ and mass bin).

Figure \ref{fig:mass-light-grad-evol} shows the running median of $\nabla\Upsilon_\star$ as a function of \tlb\ and $z$ in five bins of \ms\ (measured at the epoch of observation) and in the $g,$ $r,$ and $i$ bands. 
The associated first and third quartiles are plotted only for the $r$ band (light shaded region); the scatter corresponding to the $g$ and $i$ bands are slightly higher and lower than in the $r$ band, respectively. The dark shaded region corresponds to an estimate of the standard error of the median calculated as $\sigma_{\rm med}=1.253\sigma/\sqrt{N}$, where $\sigma$ is the standard error and $N$ is the number of elements in the given bin. The median trajectories (and the quartiles) are calculated from the highest redshift of the observed galaxies in each mass bin.
This condition ensures that when calculating the median at a given \tlb, all galaxies are included. Because more massive systems tend to have higher values of $z_{\rm obs}$, their median trajectories start with higher redshifts or look-back times.
Black dots are the { median} $t_{\rm lb,obs}$ and $\nabla\Upsilon_{\star,r}$ of the {\it observed} galaxies within the given \ms($z_{\rm obs}$) bin, with the error bars showing the respective { 25th-75th percentiles}.

According to Fig. \ref{fig:mass-light-grad-evol}, at very early epochs, the medians of $\nabla\Upsilon_{\star,r}$ are only slightly below 0 (nearly flat gradients), though the scatter is large.
On average, the more massive the galaxy, the earlier starts a period of decreasing $\nabla\Upsilon_{\star}$, which ends in a minimum.
The epochs at which the minima are reached tend to be earlier for the more massive CLE galaxies. { See also the lower right inset, where we plot the look-back time at which the (smoothed) $\nabla\Upsilon_{\star}$ trajectories reach their minima for each galaxy of mass \ms($z_{\rm obs}$). } 
After $\nabla\Upsilon_{\star}$ reached their most negative values, they increase with \tlb\ decreasing. 
In general, we observe a trend of ``delayed'' evolution as lower is the mass bin.

For CLEs less massive than $\sim 3\times 10^{10}$ \msun, the $\nabla\Upsilon_{\star}$ values can be seen to increase a little with time at very early epochs. The values of $\nabla\Upsilon_{\star}$ reach maxima around \tlb$\approx 5-7$ Gyr, after which they decrease.
For the less massive CLEs, during this maximum the gradient is even positive for a fraction of galaxies. When $\nabla\Upsilon_{\star,r}$ is positive, a more efficient SF is expected in the inner regions than in the outer ones.

Finally, as seen in Fig. \ref{fig:mass-light-grad-evol}, the scatter around the medians is large in all mass bins. This scatter is due to variations among galaxies in the given \ms\ bin (population dispersion) and due to random variations over time in the individual evolutionary $\nabla\Upsilon_{\star}$ trajectories.  Looking at the latter, we see that in most of cases $\nabla\Upsilon_{\star}$ changes randomly around their main trend over \tlb.
Probably, this is an archaeological imprint of several processes acting during the evolution of CLE galaxies: bursts of SF and quenching periods, environmental effects, and, above all, accretion of satellites of diverse masses, stellar population and gas compositions, etc. 
In section \ref{sec:scatter} we estimate the level of randomness or ``variability'' of the $\nabla\Upsilon_{\star}$ trajectories, and find that this variability contributes in quadrature $\approx 28\%$ to the dispersion around the median $\nabla\Upsilon_{\star}$ trajectories in Fig. \ref{fig:mass-light-grad-evol}. In few cases the average variability of a given trajectory may be large, of the order of the maximum variation over time of the smoothed trajectory, see \S\S\ \ref{sec:scatter} for details.

The trends seen in Fig. \ref{fig:mass-light-grad-evol} 
are very similar to those in Fig. 2 of Paper I for the \rmass/\rlight\ evolutionary tracks in the $g$, $r$, and $i$ bands. This  shows that the changes over time of the $M/L$ gradients of CLE galaxies drive the changes of their \rmass-to-\rlight\ ratios. The \rmass-to-\rlight\ ratio tell us how compact a galaxy is in mass with respect to light, and the fact that it changes with time it implies that the radial photometric evolution of galaxies differs from their radial stellar mass (intrinsic) evolution. 
In Paper I, we have discussed the physical interpretations of the inferred evolution of the \rmass-to-\rlight\ ratios of our CLE galaxies under the understanding that they explain primarily the evolution of the $M/L$ gradients reported here. We discuss these interpretation in \S\S\ \ref{sec:inside-out} below.

\begin{figure}
	\includegraphics[width=\columnwidth]{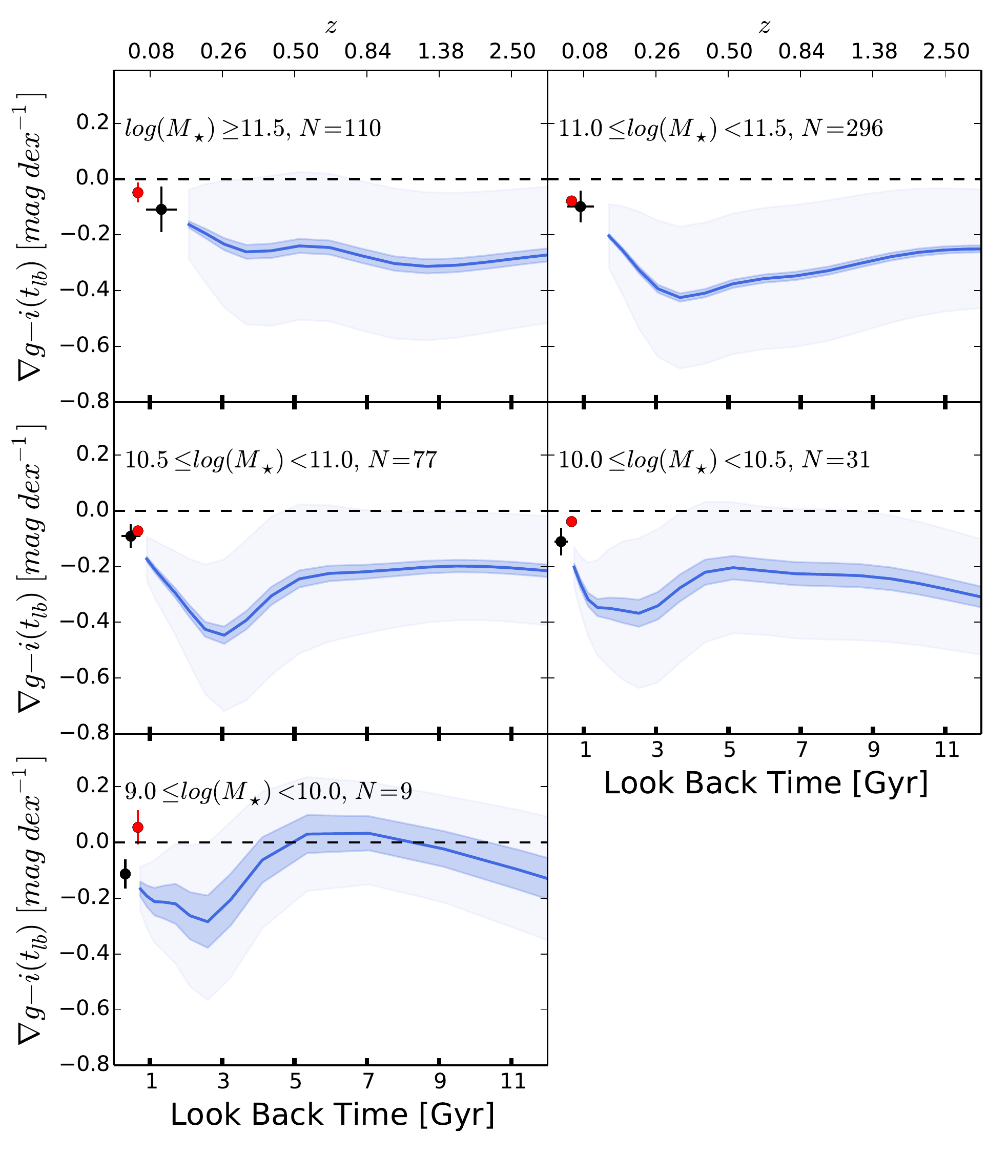}
    \caption{Evolution of $g-i$ colour gradients of our CLE galaxies. Each panel shows the running medians of the gradients in the \ms($z_{\rm obs}$)\  bins shown inside the panels (in units of \msun). The light shaded regions are the associated 25th-75th percentiles, while the dark shaded regions are estimates of the standard error of the median. 
    Black dots are the { median} $t_{\rm lb,obs}$ and $\nabla(g-i)$ of the observed galaxies within the given mass bin, with error bars showing the respective { 25th-75th percentiles}. Red dots are the average gradients in the corresponding mass bins for E galaxies from a volume-complete SDSS sample with $z_{\rm obs}<0.07$ (Stephenson et al., in prep.).
    }
    \label{fig:colour-grad-evol}
\end{figure}

Colours are quantities easier to obtain from observations than mass-to-light ratios. In principle, both are related, including their radial gradients \citep[e.g.,][]{Bell-deJong2001,Tortora+2011}. From the fossil record analysis of our MaNGA CLE galaxies, we can calculate the 2D rest-frame colour maps at any look-back time. From these maps, we measure in particular the $g-i$ gradient in a similar way as we measured the $M/L$ gradients: $\nabla(g-i)=[(g-i)^{\rm in} - (g-i)^{\rm out}]/\log(R^{\rm in}/R^{\rm out})$. Note that $\nabla(g-i)$ is in unities of mag/dex. 
Figure \ref{fig:colour-grad-evol} shows the running median and the associated first and third quartiles of $\nabla(g-i)$ as a function of \tlb\ in the same five \ms\ bins as in Fig. \ref{fig:mass-light-grad-evol}.  As expected, the trends are qualitatively similar to those of the $M/L$ gradients shown in Figure \ref{fig:mass-light-grad-evol}.  However, in more detail, the $g-i$ colour gradients at large look-back times are more negative than $\nabla\Upsilon_{\star}$ with almost not cases of positive gradients. 

The black dots in Fig. \ref{fig:colour-grad-evol} are the { median} $t_{\rm lb,obs}$ and $\nabla(g-i)$ at the observation redshift of the galaxies within the given \ms($z_{\rm obs}$) bin, with the error bars showing the respective { first and third quartiles}. The red dot in each panel is the average $\nabla(g-i)$ of E galaxies (measured within 1 \rer; Stephenson et al., in prep.) in the given mass bin taken from the $g, r,$ and $i$ 2D point spread function-corrected Sérsic fits by \citet[][]{Meert+2015,Meert+2016} of $\sim 7\times 10^5$ spectroscopically selected galaxies drawn from the SDDS DR7. The morphology has been taken from the automatic classification by \citet[][]{Huertas-Company+2011}, and the sample has been limited to $z_{}<0.07$, with a  median redshift of $z\approx 0.05$. In general, the color gradients obtained from photometric observations agree within the scatter with those calculated here from the IFS data cubes, reinforcing our results. This is not the case for the less massive bin, $9.0\le$log(\ms/\msun)$<10.0$, where the E galaxies have on average positive $(g-i)$ gradients. However, note that the E galaxies in Stephenson et al. (in prep.) are not restricted to red and quiescent as in our case.  As shown in \citet[][]{Lacerna+2020}, for less massive E galaxies, the fraction of those that are blue and/or star forming or Recently Quenched increases. The radial profiles of different properties for these E galaxies are different from the E red and quiescent ones. 
{  Also, from two-dimensional ($B-I$) color maps extending out to $\sim 2R_e$, \citet{Cibinel+2013} measured ($B-I$) color gradients around $-0.1/-0.2$ mag dex$^{-1}$ for their sub-sample of local E/S0 galaxies within the mass range $10.2\le$log(\ms/\msun)$<11.3$. These measurements are consistent with ours within the uncertainties and taking into account that our study is limited to red quiescent E galaxies.  }

\subsection{Interpreting the evolution of the $M/L$ gradient} 
\label{sec:inside-out}

As shown in Figure  \ref{fig:mass-light-grad-evol}, the $M/L$ gradients of CLE galaxies at large look-back times of 11--12 Gyr (i.e., the gradients corresponding to the observed older stellar populations) tend to be flat or slightly negative, on average. Similarly, the \rmass-to-\rlight\ ratios at these high look-back times tend to be 1 (see Fig. 2 in Paper I).  As discussed in Paper I, the above results can be interpreted as a footprint of the early rapid assembly of proto Es, when the light surface brightness profile still closely follows the shape of the stellar mass surface profile recently built-up by dissipation and strong bursts of SF. 
At those early epochs, eventual central SF bursts due to intense gas inflow (wet compaction; \citealp[e.g.,][]{DekelBurkert2014,Zolotov+2015,Tacchella+2016,Barro+2017}) can produce temporary positive $M/L$ gradients, which implies more compact distribution of light than mass. Later on, the inside-out SF quenching processes are expected to kick in, without significant structural changes in the galaxy. 
As a result, the stellar populations in the inner regions reach higher $\Upsilon_{\star}$ values than in the periphery. At the same time, the proto-ellipticals may grow externally driven by dry minor mergers, changing their structure (mainly increasing their size) and adding to the external regions most probably relatively younger and/or lower-metallicity stellar populations with lower $\Upsilon_{\star}$ values than those formed in situ in the old primary galaxy, although the satellites may contain a variety of stellar populations.
In both cases, the outcome is to make $\nabla\Upsilon_{\star}$ negative and increasingly steep (or to decrease the \rmass-to-\rlight\ ratio) as it is inferred from the fossil record analysis, see our Fig. \ref{fig:mass-light-grad-evol} and Fig. 2 from Paper I.

The steepening of the negative $M/L$ gradients continues until minima are reached at relatively late epochs. Similar to the minima of the \rmass-to-\rlight\ ratios, the minima in $\nabla\Upsilon_{\star}$ occur earlier the more massive the CLE galaxies, on average. 
In the $r$ band, the median of $\nabla\Upsilon_{\star,r}$ reaches a minimum at \tlb$\sim 2.5$ Gyr ($z\sim 0.2$) for galaxies in the $9.0\le$log(\ms/\msun)$<10.0$ bin, and at \tlb$\sim$4 Gyr ($z\sim 0.35$) for those in the log(\ms/\msun)$\ge 11.5$ bin, see Fig. \ref{fig:mass-light-grad-evol}. After the minima, the  $M/L$ gradients tend to flatten out towards lower look-back times. 
In Paper I, we have interpreted the above as consequence of the long-term global (in all radii) quenching of SF that the CLE progenitors underwent. When the external regions are quenched, their stellar populations passively age and $\Upsilon_{\star}$ increases as fast or faster than $\Upsilon_{\star}$ in the previously quenched inner regions. Therefore, the $M/L$ gradients tend to flatten slightly or, at least, remain constant. In Paper I, we have calculated the quenching look-back times of our CLE galaxies (see Appendix B there) and shown that they roughly coincide with the look-back times at which the minimum \rmass-to-\rlight\ ratios are reached (Fig. 9 in that paper), which confirms that the upturn in the evolution of this ratio is related to the global quenching. Since look-back times at which the minimum \rmass-to-\rlight\ ratios and minimum $M/L$ gradients are attained are very close, the same reasoning applies to the latter.  
{ To make sure that the minimum in the $M/L$ gradients at look-back times around 2-4 Gyr is a physical result related to CLE galaxies and not an artifact of our fossil record methodology, we have also calculated the $\nabla\Upsilon_{\star}$ evolution for MaNGA DR15 spiral star-forming galaxies. The results are presented in Appendix \ref{sec:appendix-spirals}. For most of these galaxies, their $M/L$ gradients do not attain a minimum.  
}

\begin{figure}
    \includegraphics[width=\columnwidth]{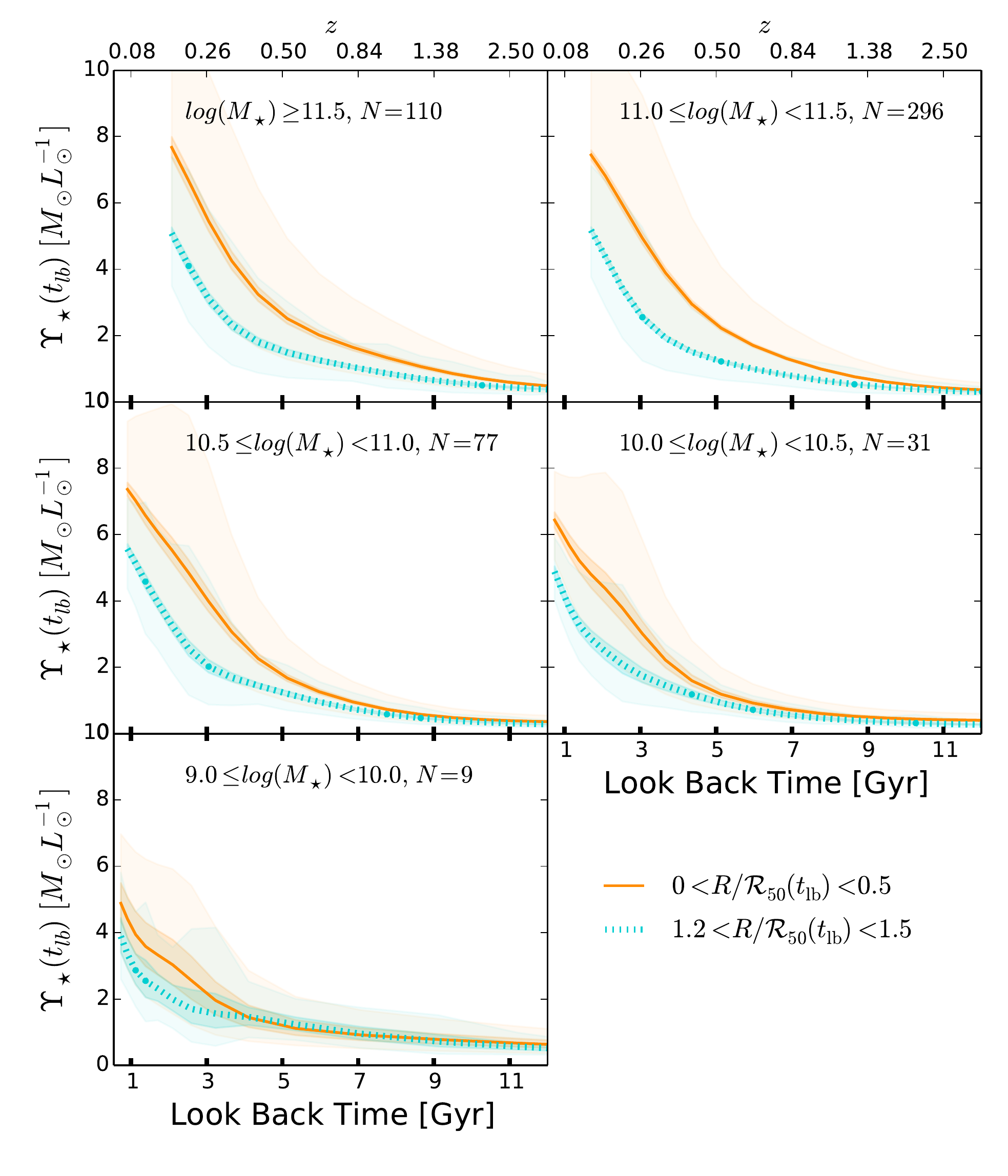}
    \caption{Evolution of the inner and outer $r-$band $M/L$ ratios. Each panel shows the respective running medians for the \ms($z_{\rm obs}$) bin indicated inside the panels (in units of \msun). The shaded regions correspond to the associated 25th-75th percentiles, while the dark shaded regions are estimates of the standard error of the median.
    }
    \label{fig:ML-evol-in-out}
\end{figure}
A relevant question is what process is more important in the evolution of the $M/L$ gradients found for CLE galaxies: inside-out quenching in a passive regime or inside-out (mainly external) mass growth with the addition of relatively younger/lower metallicity stellar populations?
In general, the global $\Upsilon_{\star,r}$ is lower than 1 \msun/L$_\odot$ at large look-back times (high redshifts) and from \tlb$\sim 8-5$ Gyr ago (the less massive the galaxy, the later the epoch), on average, $\Upsilon_{\star,r}$ rapidly increases to large values, see Fig. 5 in Paper I. 
In Fig. \ref{fig:ML-evol-in-out} here we plot the running medians of the inner and outer $M/L$ ratios in the $r-$band as a function of \tlb. These are actually the $\Upsilon_{\star, r}^{\rm in}$ and $\Upsilon_{\star, r}^{\rm out}$ used to calculate the gradients shown in Fig. \ref{fig:mass-light-grad-evol}. 
They follow a similar trend as the global $\Upsilon_{\star,r}$ but clearly $\Upsilon_{\star, r}^{\rm in}$ increases more rapidly than $\Upsilon_{\star, r}^{\rm out}$, up to late epochs when the rates become similar or even they invert. 
Stellar populations in the inner regions of CLEs appear to have aged with little or no formation of younger stars ($\Upsilon_{\star}$ correlates strongly with age, e.g. \citealp[][]{Bell-deJong2001,BruzualCharlot2003}), with this trend being greater the more massive the galaxy is. On the other hand, stellar populations in the outer regions 
also age but at a slower pace,  which may be attributed to inefficient quenching there or to the incorporation of younger/less metallic ex situ stars as part of some external growth in these galaxies. In the next section we explore what process dominates in the archaeological evolution of CLE progenitors, the inside-out quenching or the late external mass growth. 
In \S\S\ \ref{sec:caveats} we discuss caveats of our fossil record inferences due to instrumental and methodological limitations, and assumptions such as constant IMF. We show that our conclusions may change quantitatively but must remain qualitatively.

\begin{figure*}
    \includegraphics[width=2.0\columnwidth]{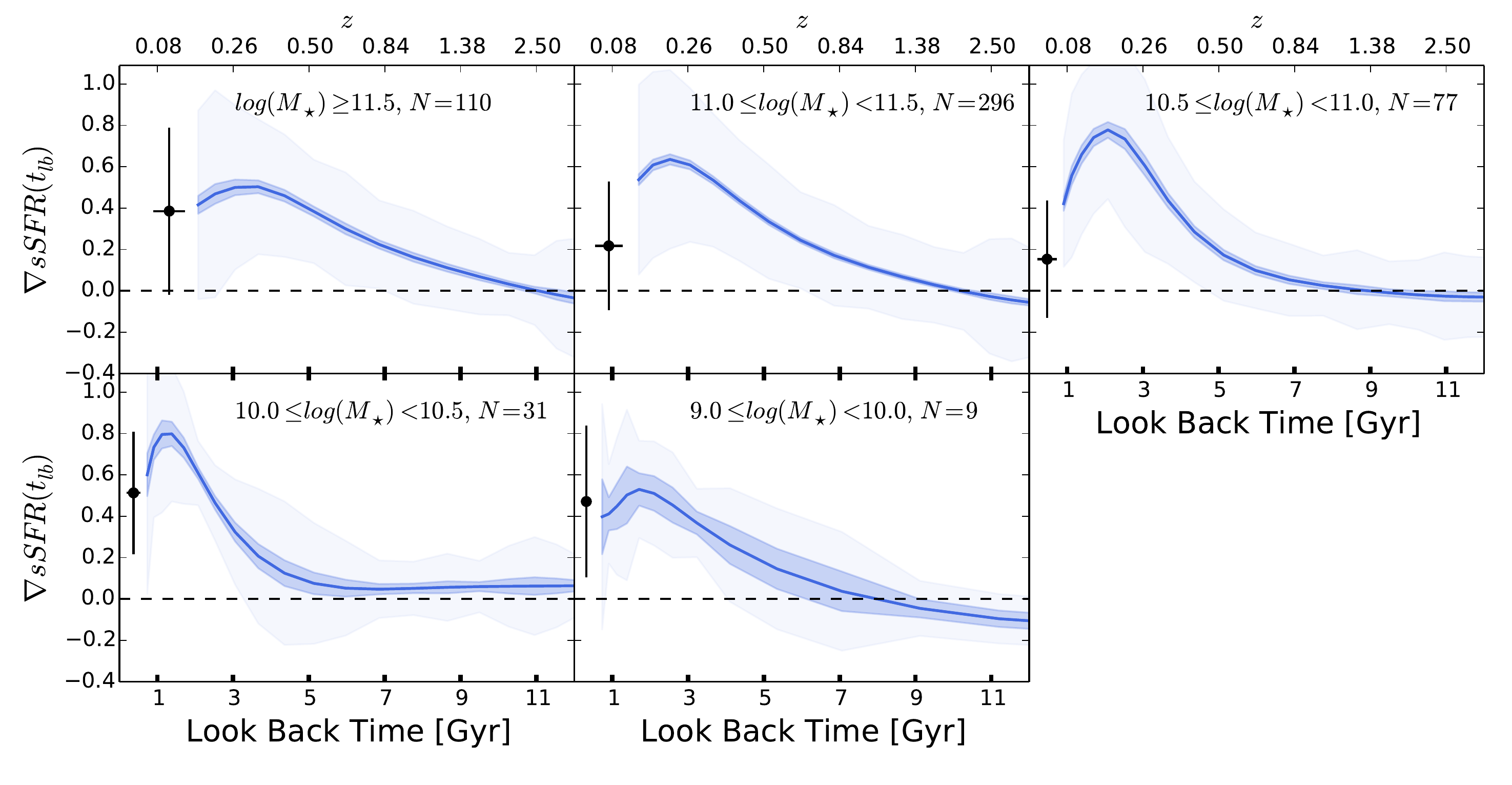}
    \caption{Evolution of sSFR gradient. Each panel shows the respective running medians for the \ms($z_{\rm obs}$) bin indicated inside the panels (in units of \msun). The light shaded regions correspond to the associated 25th-75th percentiles, while the dark shaded regions are estimates of the standard error of the median. Black dots are the { median} $t_{\rm lb,obs}$ and $\nabla sSFR$ of the observed galaxies for the given mass bin, with the error bars showing the respective { 25th-75th percentiles}.
    }
    \label{fig:sSFRgradients}
\end{figure*}

\section{Inner quenching versus outer growth}
\label{sec:quenching-vs-growth}

We have found that the $M/L$ gradient of CLE galaxies becomes more negative with time (until relatively late times, after which the gradients revert this trend),
and that this can be due to inside-out SF quenching and/or inside-out growth. Following, we explore which one dominates in the evolution of the MaNGA CLE galaxies recovered from the fossil record. We should keep in mind that the MaNGA FoVs are limited. Therefore, we will restrict our study of outer growth to radii up to 1.5\rer, the aperture of the Primary+ sample, although we also include CLE galaxies from the Secondary sample. In \S\S\ \ref{sec:caveats} we extend our study to $\approx 2.5$ \rer\ using only the CLE galaxies from the Secondary sample.

\subsection{Inside-out SF quenching}
\label{sec:in-out-quenching}

To explore the footprint of an inside-out quenching process, we have measured the inner and outer specific SF rates, $sSFR^{\rm in}$ and $sSFR^{\rm out}$, as a function of \tlb\ (see also \citealp{Lacerna+2020}), and calculate the respective gradients in sSFR in the same way as for $\nabla\Upsilon_{\star}$, see \S\S\ \ref{sec:m-l_gradients}. 
The running medians and 1st-3rd quartiles of $\nabla sSFR$ as a function of \tlb\ in the five \ms\ bins used in previous figures are shown in Fig. \ref{fig:sSFRgradients}. At large look-back times (old observed stellar populations), the median $\nabla sSFR$ is 0 (flat gradient) or slightly negative, which is typical for the regime of star-forming galaxies \citep[see e.g.,][]{Gonzalez-Delgado+2016,Sanchez+2018_AGN,Sanchez2020}. At these early epochs, the archaeological progenitors of early-type/quiescent galaxies are indeed expected to be in an active SF regime  \citep[][]{Citro+2016,Sanchez+2019,Lacerna+2020,Peterken+2021}.
At smaller look-back times, $\nabla sSFR$ becomes positive and steeper (measures of $\nabla sSFR$ indeed show positive values for local early-type galaxies, see references above), evidencing that SF quenches from the inside out. The increase in $\nabla sSFR$ occurs on average the later, the less massive the galaxies.  The positive gradients increase until reach a maximum value at look-back times that are smaller, the less massive the galaxies. For the most massive CLEs, the maximum positive sSFR gradients are reached at \tlb$\approx 3-4$ Gyr, while for the least  massive ones, the maximums are at \tlb$\sim 1$ Gyr.

On the other hand, as seen in Fig. \ref{fig:sSFRgradients}, the maximum values of $\nabla sSFR$ (when the gradients are steeper) are, on average, the larger, the less massive the galaxies (excepting for the bin of less massive galaxies, which contains only a few objects). The above is because, at epochs before the maximum gradient is reached, 
$sSFR^{\rm out}$ decreases 
slower than $sSFR^{\rm in}$ for the less massive CLEs, at least down to \ms$\sim10^{10}$ \msun. 
Regarding the scatter around the median sSFR gradient trajectories in Fig. \ref{fig:sSFRgradients}, it seems to be dominated by the diversity among the galaxies within each \ms\ bin (population dispersion). In \S\S\ \ref{sec:scatter} we show that the random variations over time in the individual evolutionary $\nabla sSFR$ trajectories contribute in quadrature $\approx 16\%$ to the dispersion seen in Fig. \ref{fig:sSFRgradients}. 
On the other hand, we see that the scatter increases, the lower the look-back time, specially in the more massive bins. This is because at these epochs, the levels of sSFR strongly decreased, as CLE progenitors quenched from the inside out. For such low levels of sSFR, the determination of sSFR and their radial gradients is very uncertain. 
This is evident from the large scatter in $\nabla sSFR$ at the observation time, as shown in Fig. \ref{fig:sSFRgradients} with the vertical error bars corresponding to the { 25th-75th percentiles} of $\nabla sSFR$ at the time of observation in each \ms\ bin, while the black dots correspond to the { median} values. As in Fig. \ref{fig:mass-light-grad-evol}, the dot also shows the { median} $t_{\rm lb,obs}$ of the observed CLE galaxies in each \ms\ bin and the horizontal error bar corresponds to the {associated  25th-75th percentiles}. 

\begin{figure}
    \includegraphics[width=\columnwidth]{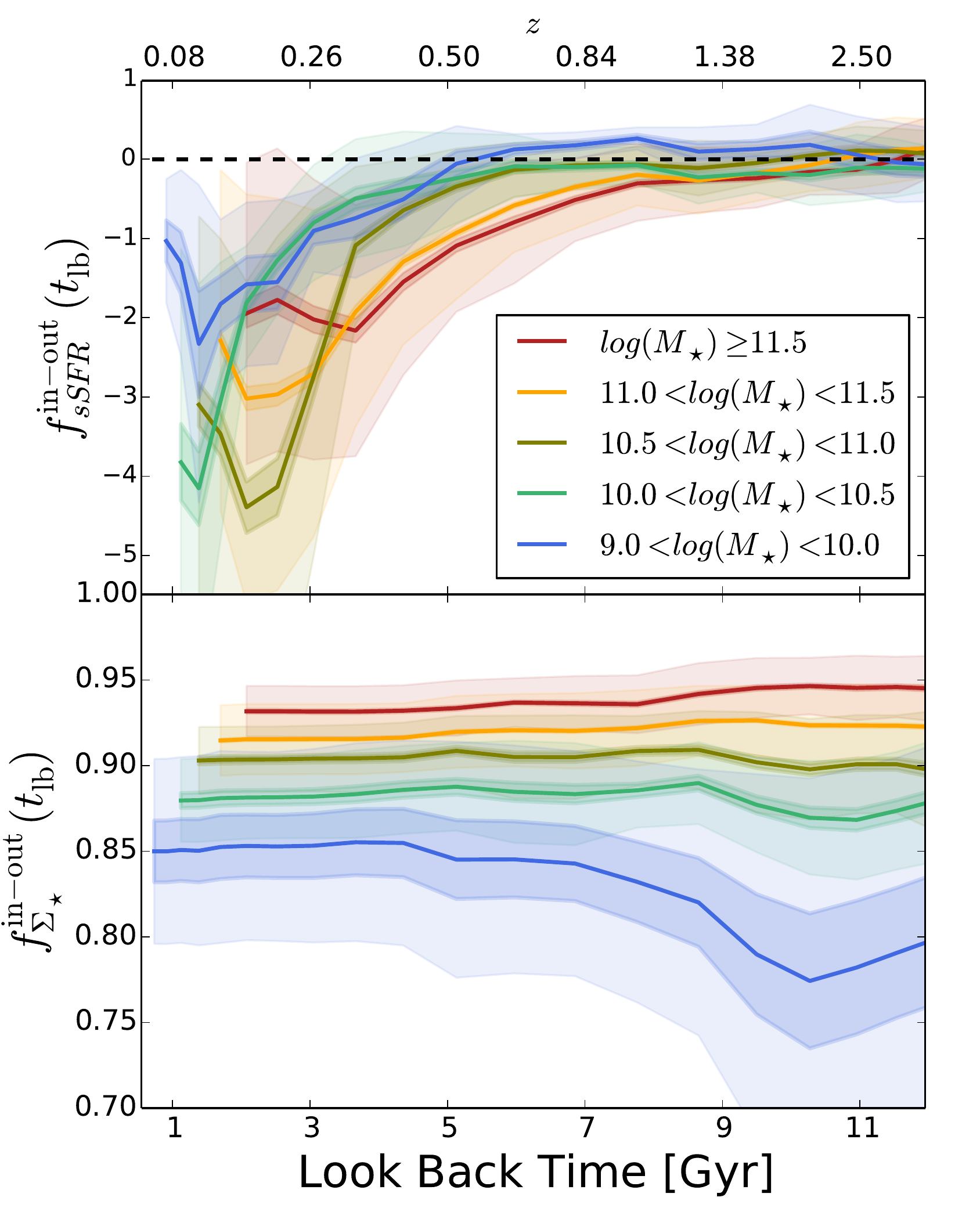}
    \caption{Change over time of the outer to inner relative differences for sSFR (upper panel) and $\Sigma_\star$ (lower panel), see Eqs. (\ref{eq:inside-outSFR}-\ref{eq:inside-out}). The selected outer and inner regions are the same at any time and are defined at 1.2-1.5\rer\ and $<0.5$ \rer, respectively.
    The above quantities tell us about the relative radial SF quenching and the relative radial stellar mass growth, respectively. Each curve shows the median trajectories over time in the five \ms($z_{\rm obs}$) bins of previous figures (in units of \msun). The shaded regions correspond to the associated 25th-75th percentiles, while the dark shaded regions are estimates of the standard error of the median. 
    }
    \label{fig:inside-out-growth}
\end{figure}

In addition to the evolution of the $\nabla sSFR$ slope of the CLE progenitors, we introduce a more quantitative metrics to assess how strong the inside-out quenching was. We compute for each galaxy, the sSFR as a function of \tlb\ within two {\it constant} in time radial intervals,  $0<R/$\rer$\le0.5$ and $1.2<R/$\rer$\le1.5$.\footnote{Note that these inner and outer sSFRs differ slightly from those used for calculating $\nabla sSFR$ in that the former are calculated at constant radii (in units of \rer) independent of \tlb.}
In fact the sSFR's are calculated along the projected semi-major axis, having in mind that $a_{e,r}=$\rer$/\sqrt{(b/a)_r}$. Then, we evaluate the change over time of the relative sSFR in the external region with respect to the internal one as follows: 
\begin{equation}
    f_{sSFR}^{\rm in-out}(t_{\rm lb})\equiv\frac{sSFR^{\rm in}(t_{\rm lb})-sSFR^{\rm out}(t_{\rm lb})}{sSFR^{\rm in}(t_{\rm lb})}.
    \label{eq:inside-outSFR}
\end{equation}
The upper panel of Figure \ref{fig:inside-out-growth} shows the running medians and 1st-3rd quartiles of $f_{sSFR}^{\rm in-out}(t_{\rm lb})$ in the five mass bins of previous figures. In early times, the sSFR in the fixed inner and outer regions were similar, on average, but then the sSFR within 0.5 \rer\ starts to decreases much faster than within the 1.2-1.5\rer\ ring. This trend starts the sooner, the higher the \ms, and it reverts in late epochs, also the sooner, the more massive the galaxy. 
The relative differences of the sSFR's seen in the outer regions with respect to the inner ones, are smaller at the upturn for higher-mass galaxies. 
CLE galaxies of intermediate masses, $10\le$log(\ms/\msun)$<11$, reach differences between the inner and outer sSFRs of factors $\sim 5$, on average, while for the most massive galaxies, log(\ms/\msun)$>11.5$, these maximum differences are of factors $\sim 3$, on average. 
The behaviors reported for $f_{sSFR}^{\rm in-out}(t_{\rm lb})$ as a function of time and mass are qualitatively similar to those of $\nabla sSFR$ shown in Fig. \ref{fig:sSFRgradients} and can be interpreted as strong inside-out SF quenching in the evolution of CLE progenitors, at least until late epochs, before the global quenching of SF occurs in these galaxies. For galaxies more massive than $10^{11.5}$ \msun, the minimum in $f_{sSFR}^{\rm in-out}(t_{\rm lb})$ happens at \tlb$\approx 3-4$ Gyr, on average, while for the less massive ones, this happens at \tlb$\approx 1$ Gyr.

\subsection{Inside-out mass growth}
\label{sec:in-out-growth}

In \citet[][]{Lacerna+2020}, for a smaller galaxy sample than here, we have shown a weak inside-out growth for the CLEs by comparing the normalized mass growth histories of their inner and outer regions. In fact, the differences between these normalized mass growth histories are much smaller than those in late-type galaxies \citep[see][]{Ibarra-Medel+2016}. Here, we quantify a better metrics for the relative mass growth of some external region with respect to another internal one, similar as we did for the sSFR in the previous section, see Eq. (\ref{eq:inside-outSFR}). By measuring the mean surface mass densities as a function of \tlb\ within two \textit{constant} in time radial intervals, $0<R/$\rer$\le0.5$ and $1.2<R/$\rer$\le1.5$ (actually in their corresponding elliptical semi-major axes), we calculate: 
\begin{equation}
f_{\Sigma_\star}^{\rm in-out}(t_{\rm lb})\equiv\frac{\Sigma_\star^{\rm in}(t_{\rm lb}) - \Sigma_\star^{\rm out}(t_{\rm lb})}{\Sigma_\star^{\rm in}(t_{\rm lb})}
 \label{eq:inside-out}
\end{equation}
The time variation of this quantity tell us the relative stellar mass growth of an external region with respect to an internal one, both regions scaled to the effective radius of the observed galaxy and kept the same over time.  
The lower panel of Fig. \ref{fig:inside-out-growth} shows the running medians and quartiles of $f_{\Sigma_\star}^{\rm in-out}(t_{\rm lb})$ in the five mass bins of previous figures. 
According to this metrics, the outer stellar mass surface density changes with time almost as the inner one, in all \ms\ bins. If any, for the two most massive bins, \ms$>10^{11}$ \msun, the median $f_{\Sigma_\star}^{\rm in-out}$ very slightly decreases with \tlb\ decreasing (differences not larger than $\sim 5\%$), which means that the stellar mass in the outer region grows slightly faster than in the inner region (inside-out growth).  
The reason that the larger \ms, the higher $f_{\Sigma_\star}^{\rm in-out}$ at any time, is due to the fact that more massive galaxies have more concentrated mass profiles. 

Our results show that the relative stellar mass growth (or the distribution of formation times of the stellar populations) of internal and external regions of CLEs have been not too different. The above is consistent with our previous finding that the age gradients of CLE galaxies are nearly flat  \citep[and the metallicity gradients are negative][see also e.g., \citealp{GonzalezDelgado+2015,Greene+2015,ZhengZheng+2017,Goddard+2017_466mass,DominguezSanchez+2019,Ferreras+2019,Parikh+2021}]{Lacerna+2020}, though these results depend on how the gradients are calculated and on how  the mean ages are defined, either mass or luminosity weighted, either arithmetically or logarithmically averaged.
To complement our analysis, for each observed galaxy, we compute in Appendix \ref{sec:appendix} the stellar mass fractions, $f_M$, contained in relatively old, intermedium, and young stellar populations both for an inner ($<0.5$ \rer) and outer (between 1.2 and 1.5 \rer) region. As expected, we find that old stellar population dominate in CLE galaxies and that their differences in fractions between inner and outer regions are very small (Fig. \ref{fig:agefrac}).

In conclusion, the fossil record analysis of our CLE galaxies shows formation time distributions for their inner stellar populations ($<0.5$\rer) roughy similar to those in a more external region (between 1.2 and 1.5 \rer), which can be interpreted, under several assumptions, as consequence of a stellar mass growth roughly homogeneous radially, at least after the dissipative phase. The above contrasts with the strong inside-out SF quenching inferred with a similar metrics and shown in the upper panel of Fig. \ref{fig:inside-out-growth}. 

\begin{figure}
    \includegraphics[width=\columnwidth]{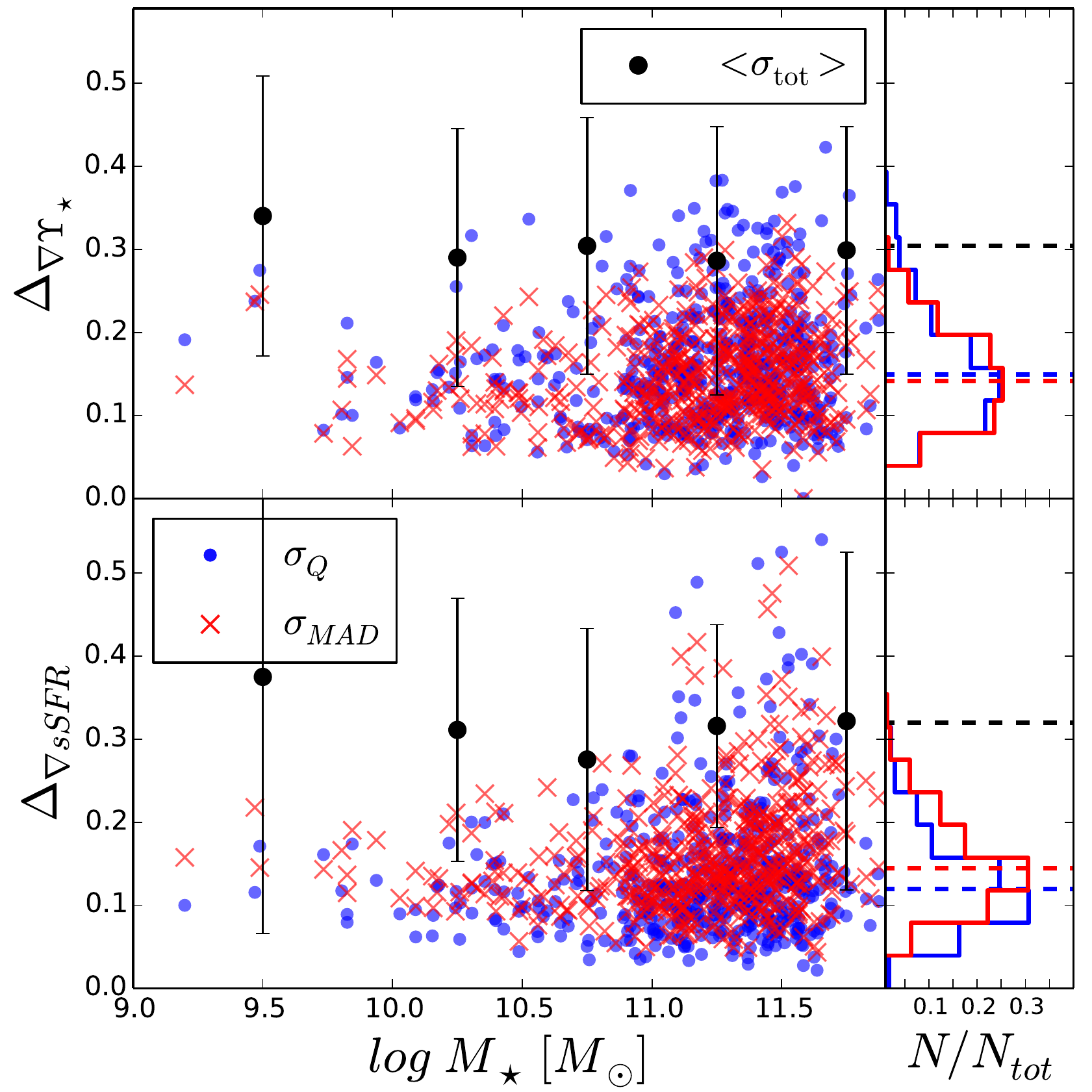}%
    \caption{Two estimates of the variability (randomness) of the individual trajectories over time of the $M/L$ (upper panel) and sSFR (lower panel) gradients as a function of \ms($z_{\rm obs}$). Blue circles are for the interquartile difference, $\sigma_{Q}$, and red crosses for the median absolute dispersion, $\sigma_{\rm MAD}$; see text. The right panels show the respective normalized distributions and their medians (dashed lines). The black dots with error bars show the total dispersion and standard deviation around the median $\nabla\Upsilon_{\star, r}$ and $\nabla sSFR$ trajectories over time in the same five \ms\ bins as in Fig. \ref{fig:mass-light-grad-evol}. The total dispersion is computed as the mean of the interquartile differences at each \tlb\ for all the galaxies within the given mass bin; see Sec. \ref{sec:scatter}.
    }
    \label{fig:scatter}
\end{figure}

\section{Individual variations versus population dispersion}
\label{sec:scatter}

The scatter around the median evolutionary trajectories of the $M/L$ and sSFR gradients shown in Figs. \ref{fig:mass-light-grad-evol} and \ref{fig:sSFRgradients}, respectively, is large. This could be due to a great diversity among individual trajectories in each mass bin (population dispersion) and/or due to random variations over time in the individual trajectories. 
To asses the variability (randomness) in a given trajectory we first smooth (average) it using a variable Gaussian kernel that considers the composed gsd156 SSP library age sampling: linear until $\approx$ 1 Gyr and then logarithmic \citep{CidFernandez+2013}. Therefore, we use a kernel with a width of 0.1 dex in logarithmic values of \tlb\ beyond 1 Gyr and with a width of 30 Myr in the linear values of \tlb. Then, we calculate the absolute residuals $\Delta_i$ at each time $t_{\rm lb,i}$ with respect to the smoothed trajectory. 
Using these residuals we compute two alternative statistical measures of global variability: (i) the median absolute dispersion, $\sigma_{\rm MAD}=$ median($\Delta_i$), and (ii) the interquartile range of the $\Delta_i$ distribution, $\sigma_{Q}= Q_3-Q_1$, where these are the 75th and 25th percentiles of the $\Delta_i$ distribution, respectively. Both quantities give an estimate of the random variations of the gradients around each smoothed gradient trajectory.  Figure \ref{fig:scatter} shows both quantities for the evolutionary trajectories of $\nabla\Upsilon_{\star, r}$ (upper panel) and  $\nabla sSFR$ (lower panel) as a function of \ms. The right panels show the respective distributions. As seen, both statistical estimates of variability in the gradients over time provide similar results. 

\begin{figure}
    \includegraphics[width=\columnwidth]{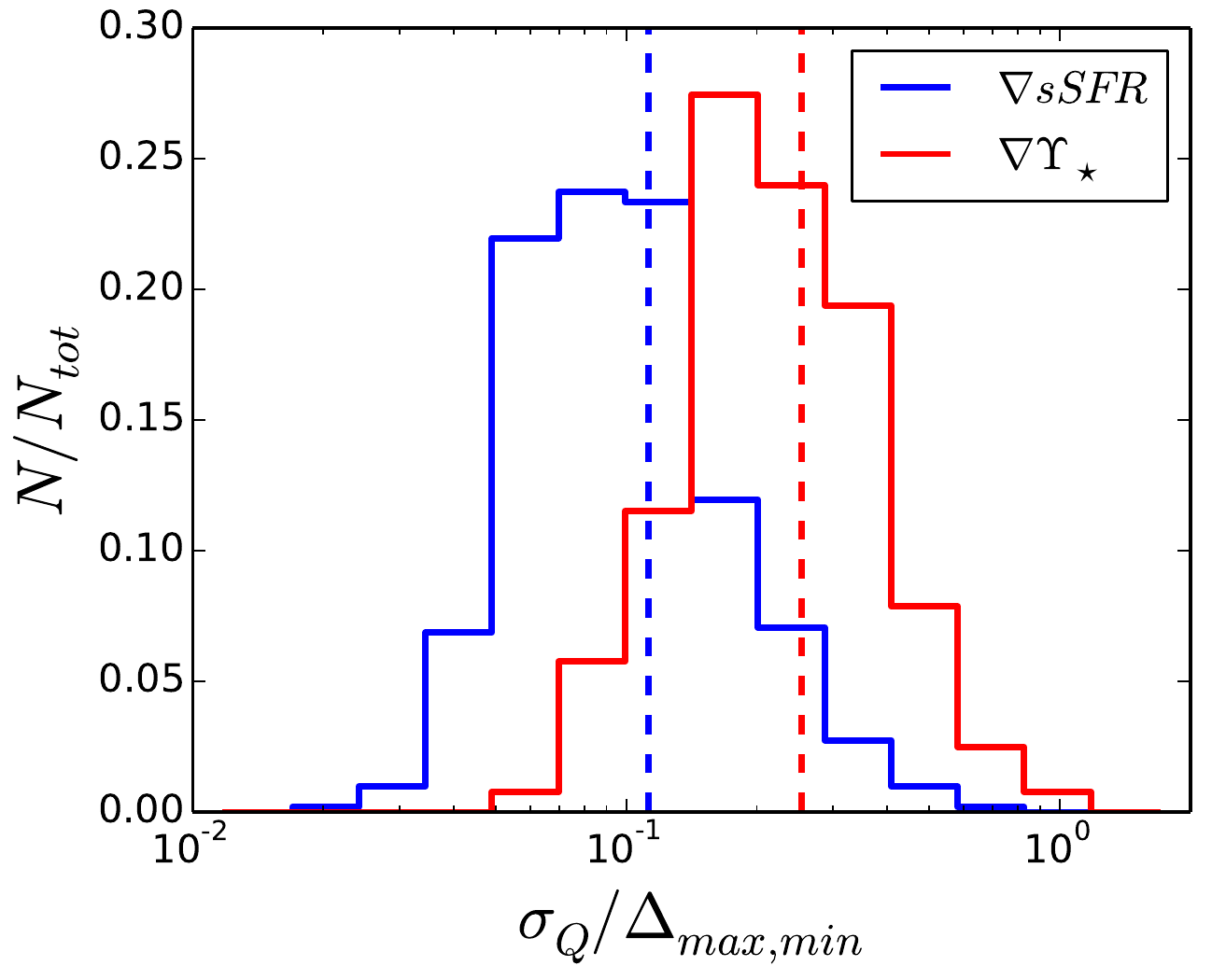}%
    \caption{Distributions of the individual variability (randomness) of the $M/L$ (red) and sSFR (blue) gradient trajectories over time (measured by $\sigma_{Q}$) relative to the maximal variation in time of these gradients when smoothed, $\Delta_{\rm max,min}$. The dashed lines indicate the median of the respective distributions.
    }
    \label{fig:scatter2}
\end{figure}

In Fig. \ref{fig:scatter} we also plot an estimate of the \textit{total} dispersion $\sigma_{\rm tot}$(\ms) of the gradient trajectories, which includes the population dispersion, for five \ms\ bins. To obtain this estimate, for a given \ms\ bin, we compute at each \tlb\ the interquartile differences $Q_3-Q_1$ of the $\nabla\Upsilon_{\star,r}$ or $\nabla sSFR$ distributions (the light shaded regions in Figs.\ref{fig:mass-light-grad-evol} and \ref{fig:sSFRgradients} show just the range between both quartiles). Then, we compute the mean and standard deviation of these interquartile differences over all look-back times, and plot them with solid black circles and error bars. The contributions in quadrature of the variability in the individual trajectories (using $\sigma_Q$) to the total dispersion around the median trajectories in Figs. \ref{fig:mass-light-grad-evol} and \ref{fig:sSFRgradients} are of $\approx 28\%$ and 16\%, respectively.
Therefore, the major contribution to the dispersion around the the $M/L$ and sSFR median gradients over time comes from differences among the individual evolutionary trajectories of the CLE galaxies in a given \ms\ bin. 
Nevertheless, the random variations around the $M/L$ and sSFR gradients are non-negligible. A way to asses how significant the random variations are is by measuring the difference between the maximum and minimum values over time of each smoothed trajectory, $\Delta_{\rm max,min}$, and compare it with our estimate of variability, for instance, $\sigma_{Q}$. 
Figure \ref{fig:scatter2} shows the distributions of the $\sigma_{Q}/\Delta_{\rm max,min}$ ratios for the $M/L$ and sSFR gradient individual trajectories. As seen, $\sigma_{Q}/\Delta_{\rm max,min}<1$ in both cases, with medians of 0.26 and 0.11, respectively. Therefore, the variations around the mean (smoothed) trajectories are significantly lower than their maximal differences over time for the majority of CLE galaxies. In the case of the $M/L$ gradient, $\approx 11\%$ of CLE galaxies have relatively high random variations, such that $\sigma_{Q}>0.5\Delta_{\rm max,min}$.

\section{Discussion}

\subsection{Caveats}
\label{sec:caveats}

 The spatially-resolved analysis with the fossil record method applied to the MaNGA galaxies is subject to many uncertainties, degeneracies, and limitations, several of which were tested \citep[see e.g.,][Paper I, and more references therein]{Sanchez+2016_p21,Sanchez+2016_p171,Ibarra-Medel+2019,Sanchez2020}. These caveats are related to: 
(a) instrumental aspects, such as the the number of fibers in the bundles (therefore, the spatial sampling/resolution) and the aperture covering; (b) observational settings, such as the galaxy inclination and the SNR; and
(c) systematic and random uncertainties intrinsic to the fossil record method (degeneracy of the spectral features for old populations, age-metallicity degeneracy, assumptions about the IMF, use of different stellar libraries, etc.). 
On the other hand, the fossil record reconstruction refers to stellar populations at their (spatially resolved) positions in the observed galaxy, but they could have been formed {in other galaxy places and then migrated or even in other galaxies (ex situ) and aggregated through dry mergers.}

{ There is an extensive amount of literature about the uncertainties of the fossil record method \citep[e.g.,][]{Ocvirk+2006,Conroy2013,Ge+2018}. In particular, in the case of the \verb|Pipe3D| code and the SSP template used here, many tests were performed to explore how much these uncertainties affect the recovered SSP mix, and therefore, the inferred star formation and metallicity histories \citep[see e.g.,][]{Sanchez+2016_p21,Sanchez+2016_p171,Mejia-Narvaez+2022}. For instance, \citet{Sanchez+2016_p21} studied several template libraries with different age and metallicity discrete samplings and have found that the age-metallicity degeneracy is weaker when using the gsd156 template library and a SNR$>50$ across the spectrum. This is one of the reasons \verb|Pipe3D| uses this template and it performs the segmentation in order to achieve an integrated SNR$>50$. In \citet[][]{Ibarra-Medel+2019}, we also tested the Pipe3D code against mock datacubes created using post-processed hydrodynamical zoom-in cosmological simulations (presented in \citealp{Colin+2016}  and \citealp{Avila-Reese+2018}), 
and have shown that the ability of the code to recover archaeological global/radial SF and stellar mass growth histories is reasonably, especially for stellar populations corresponding to early-type galaxies. The SPS methodology introduces a certain level of uncertainty that it cannot avoid, but we can estimate it through all the tests that we have previously published. As we show in \citet[][see also \citealp{Sarmiento+2022}]{Ibarra-Medel+2019}, the main sources of uncertainty actually are not  those from the SPS code, but rather are due to the observational setting and instrumental limitations (e.g., inclination, spatial resolution/number of fibers, SNR, etc.).
}

In Paper I \citep[section 5.1; see also][]{Lacerna+2020}, we have discussed extensively all the caveats mentioned above and how they could affect the global and radial archaeological inferences for the same MaNGA CLE galaxies studied here. We refer the reader to these papers for details. 
The main effects of the mentioned uncertainties, degeneracies, and limitations of the method for quiescent early-type galaxies are to smooth the SF and mass growth histories for ages older than $\sim 7$ Gyr and to flatten the inferred radial gradients of the stellar population properties. Therefore, the radial trends and their evolution presented here could actually be {\it even more pronounced}, but they are expected to remain at a qualitatively and statistical level.

\subsubsection{Spatial resolution and PSF effects}
\label{sec:spatial-resolution}

Using mock observations of post-processed simulated galaxies, \citet[][]{Ibarra-Medel+2019} have shown that the accuracy in the recovery of the radial variations of the stellar population properties worsens mainly as poorer the spatial resolution and SNR are, and more inclined is the galaxy. The latter is not too worrying for E galaxies. As done in Paper I, in order to asses statistically how much our result can be affected by fiber sampling (spatial resolution), we have calculated again the running medians of $\nabla\Upsilon_{\star,r}$ presented in Fig. \ref{fig:mass-light-grad-evol}
but only for those galaxies best spatially resolved; for the way the MaNGA survey was designed, they correspond to galaxies sampled with bundles containing 127 and 91 fibres. The above condition reduces the sample to $\sim 25$ per cent. Figure \ref{fig:HR-mass-light-grad-evol} shows the results for the best spatially resolved galaxies, which can be compared with those for the whole CLE sample plotted in Fig. \ref{fig:mass-light-grad-evol}. 
We do not present results for the two low-mass bins because there are only a few observed CLE galaxies with 127 and 91 fibres (5 in total). As expected, the corresponding median $\Upsilon_{\star,r}$ gradients are steeper for most of the look-back times (and specially at the minimum) than for the whole sample. 
Similar results are obtained for the medians of the colour and sSFR gradient tracks presented in Figs. \ref{fig:colour-grad-evol} and \ref{fig:sSFRgradients}.  The above shows that limitations in the fiber sampling (spatial resolution) lead to underestimating the $M/L$, colour and sSFR gradients, as expected, but the underestimations are moderate and the qualitative trends do not change. { Moreover, part of the differences in the $M/L$ gradients mentioned above can be rather intrinsic in the sense that galaxies observed with 91 and 127 fibers, by construction of the sample, are biased to be more extended at a given mass than the average \citep[][]{Wake+2017}. More extended or less compact galaxies tend to have steeper gradients. }

\begin{figure}
    \includegraphics[width=\columnwidth]{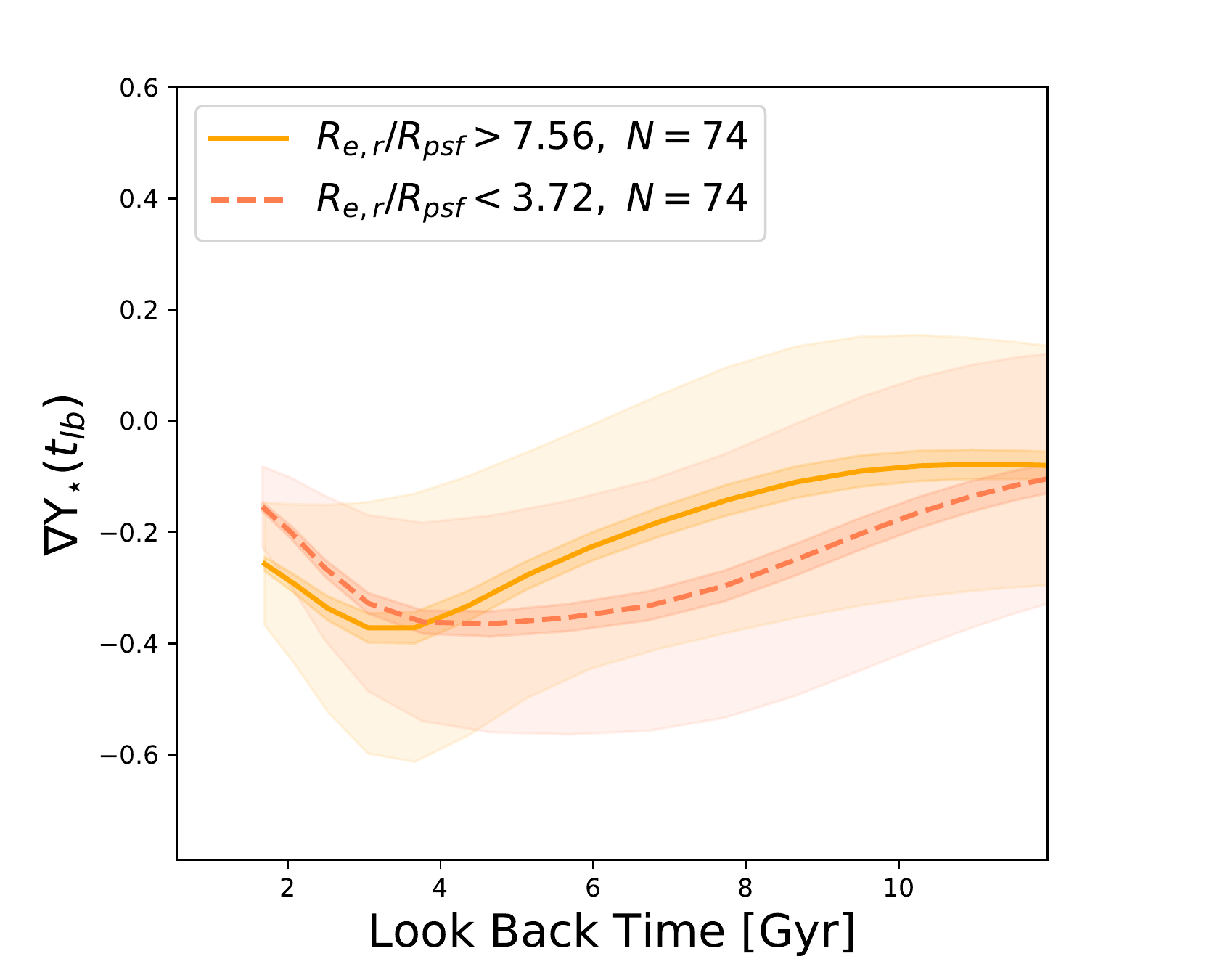}
    \caption{
    { Evolution of the M/L gradient of the MaNGA CLEs within the observed mass bin of 11$<$log(\ms/\msun)$\le 11.5$. The solid orange line is for galaxies above the first quartile of the \rer/$R_{\rm PSF}$ distribution within this mass bin, 
    while the red segmented line is for those below the third quartile.
    The light shaded regions correspond to the respective 25th-75th percentiles, while the dark shaded regions are estimates of the standard error of the median.}
    }
    \label{fig:high-low-resolution}
\end{figure}

{  The MaNGA PSF is one of the factors that can affect the inferred radial gradients \citep[see e.g.,][]{Ibarra-Medel+2019}. This is not a problem when the apparent galaxy size is much larger than the PSF, especially when using large-scale (global) gradients like the ones we measured, avoiding more local determinations. For most of the CLE MaNGA galaxies studied here, \rer\ is much larger than the PSF radius, $R_{\rm PSF}$. The median of \rer/$R_{\rm PSF}$ is 5.4. As mentioned in \S\S\ \ref{sec:data}, the galaxy sample studied here includes only galaxies with  \rer/$R_{\rm PSF}>1.75$ to minimize any effect of the PSF on our results. 
In any case, to explore the effect of low spatial resolution in terms of the PSF on the inferred $M/L$ gradients, we calculated the median evolutionary tracks by separating galaxies into the 25\% with the largest \rer/$R_{\rm PSF}$ ratios and the 25\% with the smallest ratios. Figure \ref{fig:high-low-resolution} shows the results for galaxies in the 11$<$log(\ms/\msun)$\le 11.5$ bin, which is the most populated. The general trend is the same in both cases. 
In more detail, although there are not significant differences within the 25th-75th percentiles, the galaxies that could be more affected by the PSF (observed with less spatial resolution) tend to have, on average, slightly flatter $M/L$ gradients at late epochs but steeper at earlier epochs than those galaxies better sampled in terms of the PSF. 
 
In summary, the spatial resolution in terms of the size of the PSF slightly affects the measurement of radial gradients but not to the point of making our main conclusions invalid or uncertain.}

\begin{figure}
	\includegraphics[width=\columnwidth]{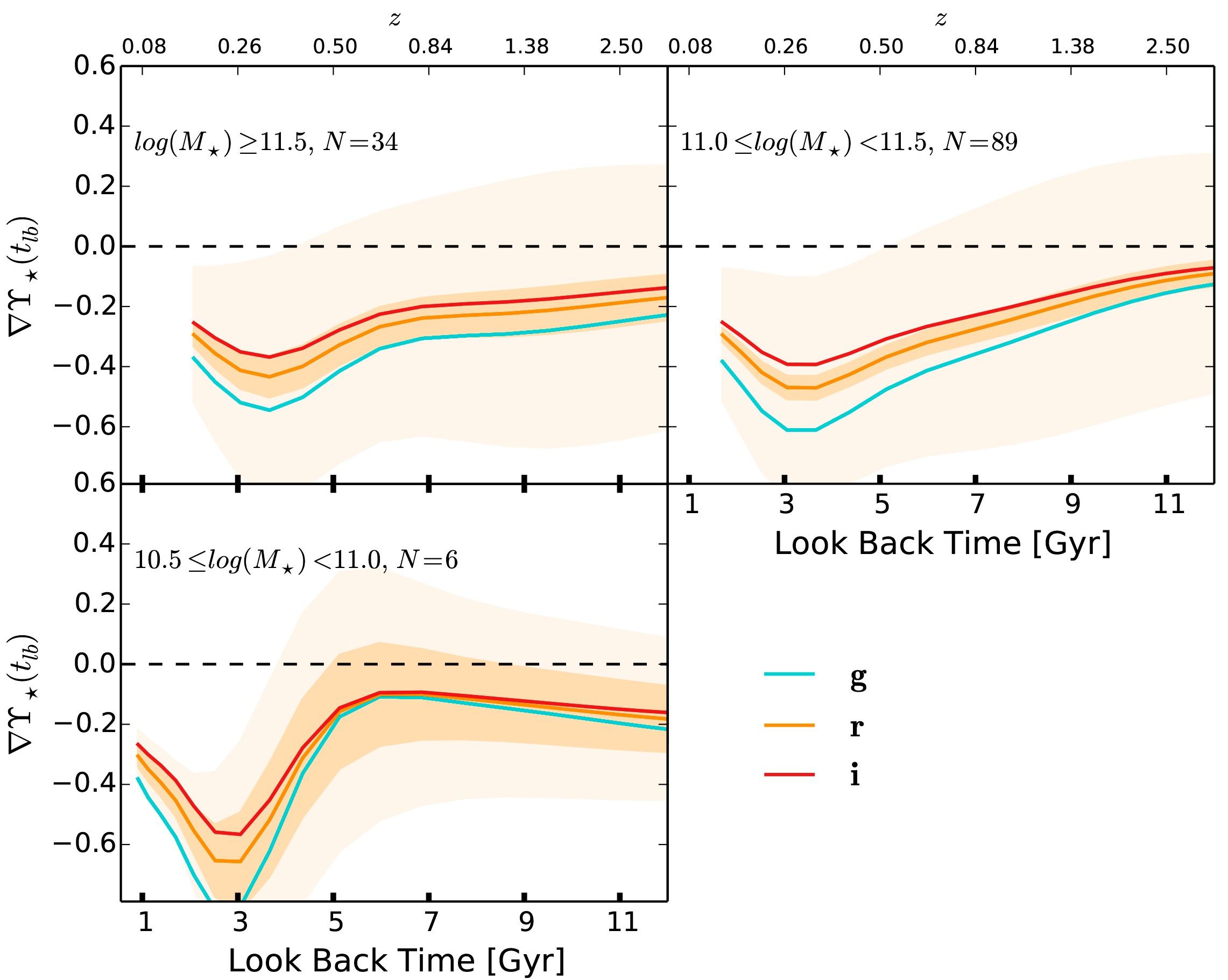}
    \caption{As Fig. \ref{fig:mass-light-grad-evol} but for only well spatially resolved CLE galaxies, i.e., those sampled with 91 and 127 fibers ($\sim 25$ per cent of the sample). The results for the two low-mass bins are not shown because there are only a few galaxies in them.
    }
    \label{fig:HR-mass-light-grad-evol}
\end{figure}

{
\subsubsection{Stellar migration and dry mergers}

In \S\S\ 5.1.2 of Paper I  we have discussed the effects of stellar migration on the radial archaeological inferences \citep[see also,][]{Ibarra-Medel+2016}. Several studies have shown that  the net radial migration is mostly outwards and only significant in the outermost parts of galaxies (see references in Paper I), so likely to be beyond the FoV of the MaNGA observations.   
On the other hand, radial displacements from one side to another less than 1--3 kpc are not actually relevant to the inferences obtained for MaNGA galaxies because these scales are not resolved. Regarding dry mergers, which are expected to affect more the evolution of E galaxies later in time, their net effect goes in the direction of flattening radial differences in the stellar populations of the main progenitor, though the final outcome depends on many pre-merger conditions and the number of mergers \citep[e.g.,][]{DiMatteo+2009}.
On the other hand, note that the radial gradients we measure in this paper are calculated over a very wide radial range such that they are not too affected by relatively local dynamical processes.

In summary, both stellar migration and dry mergers work in the direction of flattening any radial gradient in the properties of stellar populations within the main progenitor. Therefore, most archaeological inferences will tend to show somewhat flatter radial distributions than the actual progenitor might have had  in case of migrations and/or stellar mergers. Only in the extreme cases of recent major dry mergers, archaeological inferences of the evolution of radial gradients at early (pre-merger) times could be misleading.
}

\subsubsection{Variations in the IMF within galaxies}

A question of concern in our archaeological inferences is that we assume an universal IMF across and among galaxies.  As discussed in Paper I, there are some pieces of evidence that early-type galaxies have different IMFs in their internal and external regions, such as a Salpeter one or heavier in the core and bottom-light IMFs (such as the Chabrier one) at larger galactocentric radii; see the references therein and \citet[][]{Smith2020} for a recent review. Furthermore, it appears that for early-type galaxies, the \textit{local} IMF is tightly related to the \textit{local} metallicity \citep[][and more references therein, { but see \citealp{LaBarbera+2019}}]{MartinNavarro+2015}, so that a negative metallicity gradient may imply the IMF variation mentioned above.
The spectral inversion method for archaeological inferences is complex and relaxing the assumption of an universal IMF within galaxies introduces strong uncertainties in the results. On the other hand, the use of line-strength indices, specially TiO$_2$, while it provides valuable clues to the IMF and $M/L$ gradients, require high signal-to-noise ratio (SNR) spectra and may be affected by several degeneracies and uncertainties. 
\citet[][see also \citealp{Bernardi+2022}]{DominguezSanchez+2019} have applied this technique to stacked spectra of early-type MaNGA galaxies and found the above mentioned trends for the IMF with radius.  

Based on the results by  \citet[][see their figures 16--18]{DominguezSanchez+2019}, {\it we expect then that the variations in the IMF produce $M/L$ gradients steeper than when the IMF is constant}. Therefore, in general our $M/L$ gradients could be underestimated, so they can be considered as lower limits. Furthermore, according to \citet[][]{DominguezSanchez+2019}, the main changes in our $M/L$ ratios, in case of allowing for a variable IMF, are expected to happen in the outer regions, that is for $\Upsilon_{\star}^{\rm out}$ (should be lower given that the IMF there tend to be Chabrier), while in the inner regions the IMF is close to Salpeter, the one we assume here, so that the $\Upsilon_{\star}^{\rm in}$ values are expected to remain similar as in Fig. \ref{fig:ML-evol-in-out}.
In addition, given that older populations are those thought to have more bottom-heavy IMFs (close to the Salpeter IMF), we expect the $M/L$ gradients archaeologically determined in Fig. \ref{fig:mass-light-grad-evol} at high look-back times (old populations today) 
to be less affected than at low look-back times, when the fraction of young populations increases. 
On the other hand, according to these and other authors, the lower the mass of early-type galaxies, the more constant the IMF with radius (this is also consistent with their flatter metallicity gradient as the mass is lower, and taking into account that the local IMF correlates with the local metallicity), so the potential effect of a varying IMF within the galaxy on the $M/L$ gradient is expected to be important only for our massive CLE galaxies. { However, even for these galaxies, the strongest evidence for substantial IMF variation only corresponds to their cores \citep[e.g.,][]{vanDokkum+2017,LaBarbera+2019,Smith2020}.}

In summary, the results presented here for the evolution of the $M/L$ gradients of CLE galaxies are expected to remain qualitatively the same if relaxing the assumption of constant IMF within the galaxies, { something that seems to apply only to the cores of the most massive E galaxies. }


\begin{figure*}
\includegraphics[width=\columnwidth]{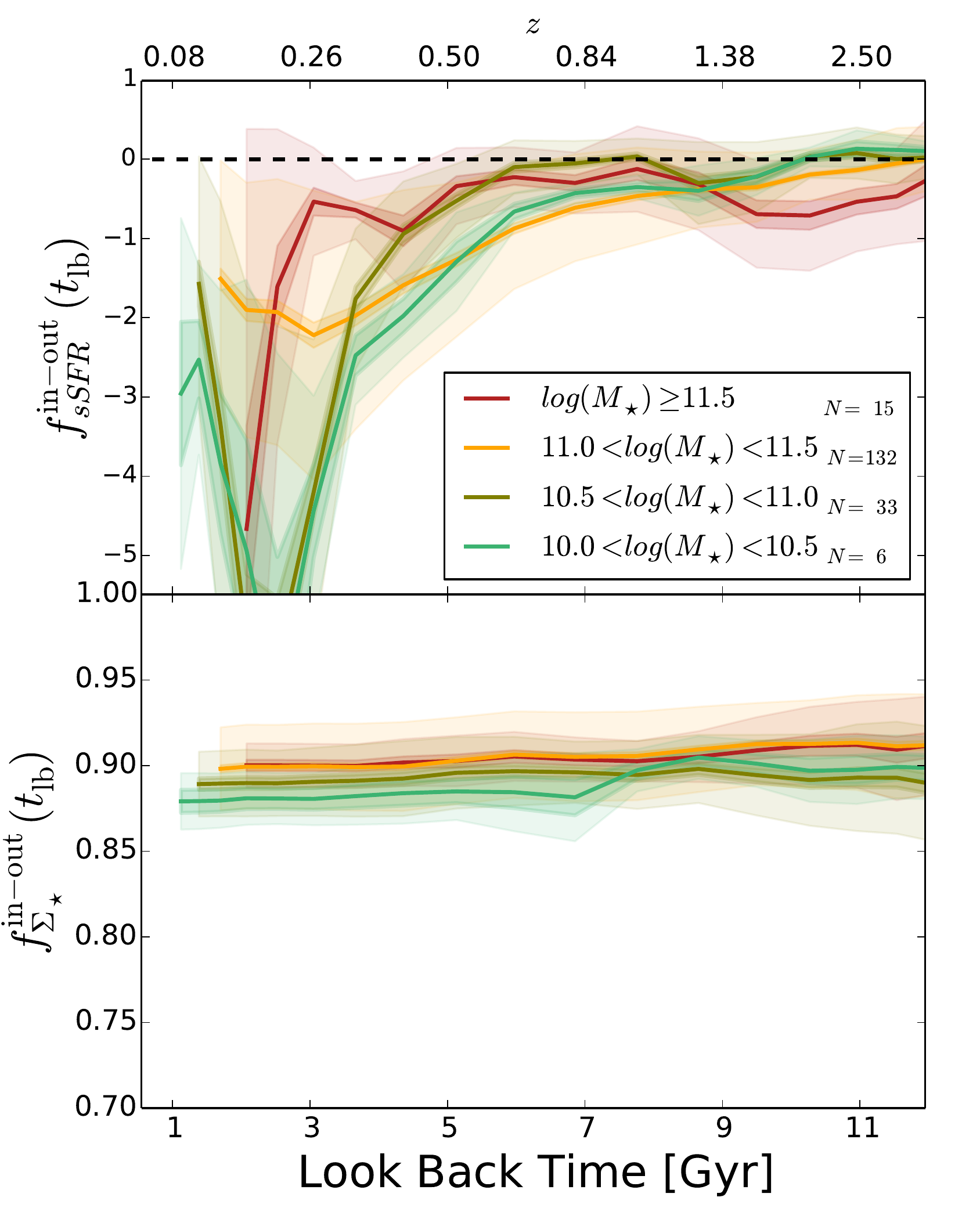}
 \includegraphics[width=\columnwidth]{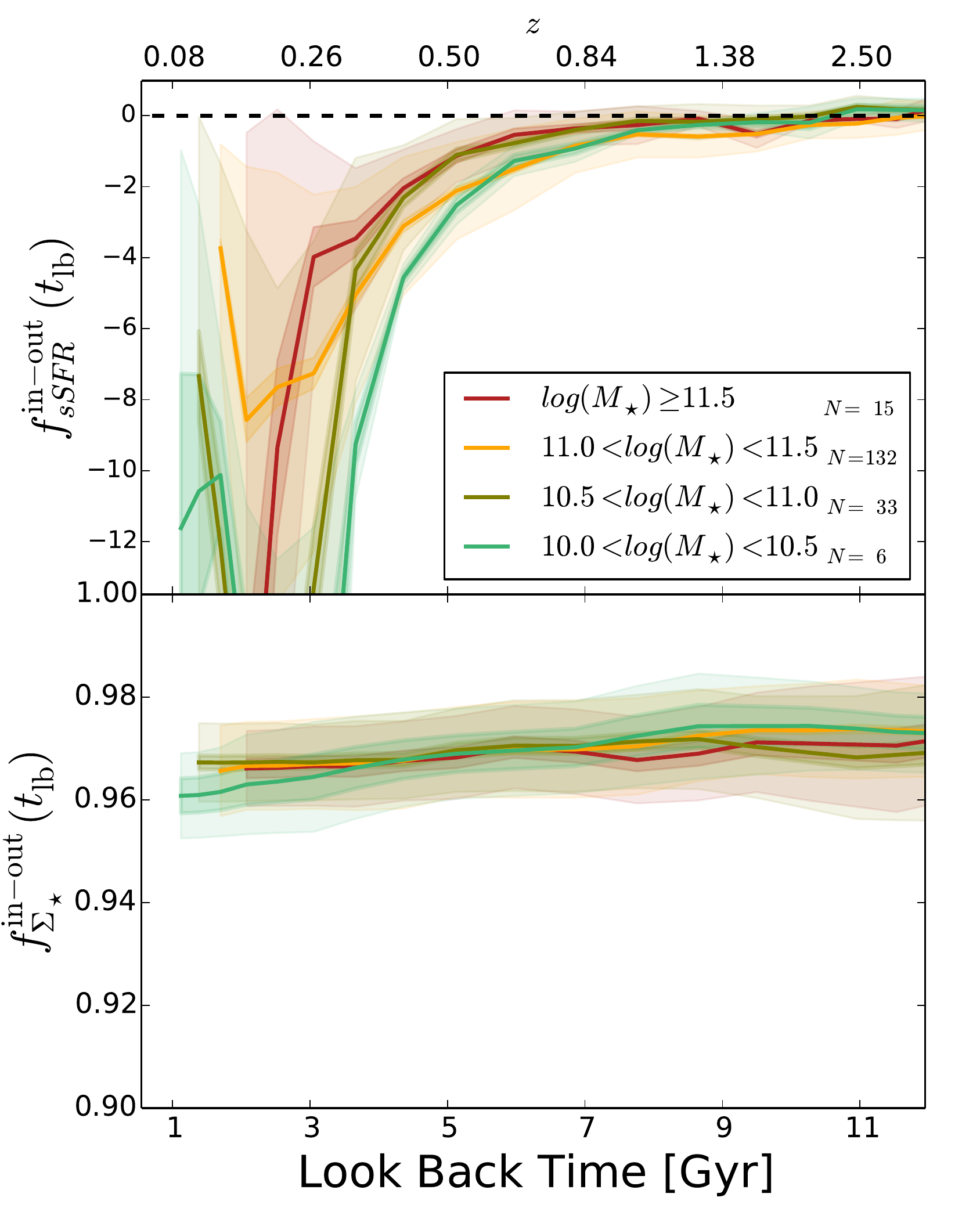}
    \caption{{ Left panels:} As Fig. \ref{fig:inside-out-growth} but for only CLE galaxies from the MaNGA Secondary sample. In this case, there are no CLE galaxies in the lowest mass bin. { Right panels: Same as left panels but} using for the outer region 2.0-2.5\rer\ instead of 1.2-1.5\rer. Note that the ranges on the vertical axes are different from those in the left panels.
    }
    \label{fig:inside-out-growth-secondary}
\end{figure*}

\subsubsection{Aperture limits}
The radial extension considered here to study the external sSFR and stellar mass histories (as well as the different gradients) is up to 1.5 \rer, the aperture of most of the galaxies observed in MaNGA. Our conclusion that the external and internal regions of observed CLE galaxies had roughly similar mass growth rates after their dissipative phase (\S\S\ \ref{sec:in-out-growth}) could differ for regions more external than 1.5 \rer. For example, \citet[][]{Oyarzun+2019}, analysing the stellar metallicity radial profiles of MaNGA passive early-type galaxies find that these profiles flattens at $\gtrsim 2$ \rer\ for galaxies more massive that $\sim 10^{11}$ \msun, and, { based on the results of hydrodynamical simulations \citep[e.g.,][]{Cook+2016},} they attribute this to stellar accretion. Using a toy model for metallicity profiles, they estimated that ex situ stars at radii $\gtrsim 2$ \rer, that is, stars accreted lately in dry mergers, make up large fractions of the total stellar mass of the most massive galaxies, supporting the inside-out mass growth scenario.

In Fig. \ref{fig:inside-out-growth-secondary}, { left panels, }we repeat Fig. \ref{fig:inside-out-growth} but only for CLE galaxies from the MaNGA Secondary sample, that is, those observed with apertures covering $\sim 2.5$ \rer\ ($\approx 36\%$ out of the CLE sample). { In the right panels,} the external region was taken in a ring between 2 and 2.5 \rer, 
that is, we extended our exploration to more external regions. 
As seen in the lower panels, even extending the outer regions to $2.0<$\rer$\le2.5$, we do not find evidence of a different archaeological stellar mass growth of these regions with respect to the inner ones. Regarding the sSFRs, the differences between the outer and inner regions change dramatically over time when shifting the outer region from $1.2-1.5$\rer\ to $2.0-2.5$\rer. This shows that SF quenches less efficiently, the larger the radius is, and confirms the strong inside-out SF quenching that CLE progenitors underwent, specially at $z<1$.

\begin{figure}
    \includegraphics[width=\columnwidth]{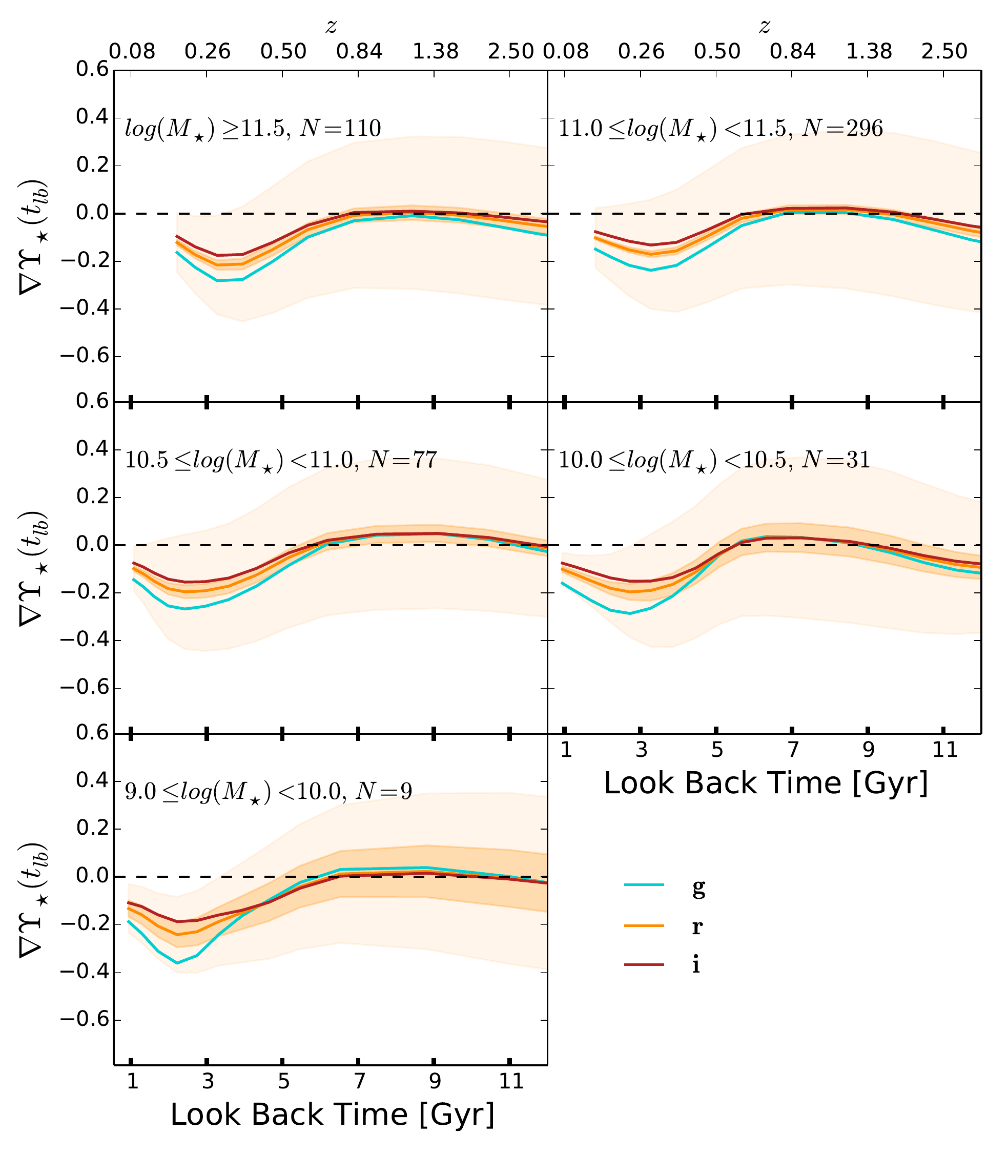}
    \caption{As Fig. \ref{fig:mass-light-grad-evol} but using the MaStars stellar library.
    }
    \label{fig:MaStars-MLgrad-evol}
\end{figure}
\subsubsection{Changing the stellar library}
A major concern in spectral inversion methods based on stellar population synthesis is the dependence of the results on the stellar templates used for the SSPs. 
\citet[][]{Sanchez+2016_p21,Sanchez+2016_p171} have shown that the \verb|Pipe3D| code recovers roughly similar properties of the stellar populations using different SSP stellar libraries for the templates, including the one adopted here (see Section \ref{sec:data}). Moreover, the retired ellipticals are expected to be the least affected by implementing other libraries and SSP age sampling as their stellar populations, on one hand, have typically metallicities higher than the limit allowed by the gsd156 library used here, and in the other hand, they are older and more homogeneous than those of late-type galaxies. 
In any case, we have analized also the results for our CLE galaxies employing an update of the \verb|Pipe3D| code, the \verb|pyPipe3D| code \citep[][]{Lacerda+2022,Sanchez+2022}, which uses the MaStars stellar library \citep{Yan+2019,Mejia-Narvaez+2022} and the GALAXEV code \citep[][]{BruzualCharlot2003} to generate a wide set of synthetic SSP templates.

Figure \ref{fig:MaStars-MLgrad-evol} compares the running medians of $\nabla\Upsilon_{\star,r}$ showed in Fig. \ref{fig:mass-light-grad-evol} with those obtained using the MaStars library and GALAXEV code. Trends are the same qualitatively, but at the quantitative level there are differences. Our archaeological inferences with the MaStars library with respect to the gsd156 imply shallower $M/L$ gradients in general, and a later start of the negative steepening of gradients in the case of massive galaxies. The gradients at look-back times $\gtrsim 6-7$ Gyr ago were on average flat with a significant fraction being even positive, which as mentioned above is typical of star-forming galaxies in a strong regime of central SF. We have noticed that using the MaStars library, the mass-weighted ages recovered for the oldest populations are always smaller than when the gsd156 library is used. Consequently, the $M/L$ ratios can not reach values as high as in the case of the gsd156 library. The above implies that, when the MaStars library is used, the $M/L$ ratios in the inner regions (where the highest values are reached) of our CLE galaxies are closer to those in the outer regions, and therefore the $M/L$ gradients are flatter. In general, for the MaStars library, the distribution of the highest $M/L$ ratios and oldest mass-weighted ages move to lower values with respect to those obtained when the gsd156 library is used.  
This can be partially due to the new range in metallicities covered by the SSPs in the MaStars library. 

\subsection{Implications of our results}
\label{sec:implications}

The spatially resolved fossil record analysis applied to the MaNGA subsample of red and quiescent E galaxies shows systematical changes in their $M/L$ and colour gradients over time or $z$, \S\S\ \ref{sec:m-l_gradients}. It is expected that many limitations and uncertainties in this kind of inferences affect our results, but not as to change the key conclusions, as discussed in the previous subsection. 

In \S\S\ \ref{sec:inside-out} we have discussed interpretations to our results in the light of evolutionary processes proposed for the progenitors of present-day CLE galaxies, and in Section \ref{sec:quenching-vs-growth} we have explored fossil record footprints of inside-out SF quenching versus external mass growth. 
Our results support a scenario in which the progenitors of CLE galaxies, after an intense and fast dissipative phase, evolved in a quasi-passive regime driven mostly by a gradual steepening of their negative $M/L$ gradient. In this quasi-passive regime dominates the process of inside-out SF quenching until the galaxy quenches globally and definitively. 
The more massive the CLE galaxy, the earlier starts its inside-out quenching and the earlier quenches globally, on average. After the global quenching of SF, the $M/L$ gradients revert their steepening, as the light radial distribution passively tends to that of the stellar mass.

Our archaeological inferences have some implications  for interpreting the evolution of progenitors of present-day quiescent E galaxies, summarised in the following.

\begin{itemize}
    \item { Although the fossil record reconstruction of the radial evolution of galaxies does not allow us to trace their dynamical history, from the results and discussion presented in this paper (see especially Section \ref{sec:quenching-vs-growth}), we can propose that the progenitors of the studied CLE galaxies} instead of a { non-dissipative} strong structural evolution, seem to have undergone radial photometric changes. The latter led to systematic decrease (steepening) of their $M/L$ gradients, { at least within $\sim 1.5-2.0$ \rer}. The main driver of such changes { seems to be}  a process of SF quenching from inside out. However, in some cases, external mass growth, {  indicative of minor mergers,} can also be relevant.
    
    \item As a consequence of the photometric evolution in a quasi-passive regime, the \rmass-to-\rlight\ ratio tends  to decrease over time (Paper I). Therefore, the strong size evolution of quiescent or early-type galaxies found in many photometric studies of galaxy samples at different redshifts could be partially consequence of the more negative $M/L$ gradients that these galaxies have as $z$ is lower, as has recently been shown by \citet[][]{Suess+2019a,Suess+2019b} and \citet[][]{Miller+2022}. 
    
    \item The progenitors of local CLE galaxies were in the past actively forming stars \citep[][see also Paper I]{Lacerna+2020}. Their (relatively late) global shut-down of SF probably happened when the galaxy has already undergone its morphological transformation. The latter is supported by the lack of archaeological signatures of significant changes in the radial stellar mass distribution since high look-back times, while the sSFR strongly changes during this time globally and radially. The above suggests that among early-type (bulge-dominated) galaxies, the probability to find that they have SF activity (or are blue) increases with $z$ (the recent results by \citealp[][]{Dimauro+2022}, who performed bulge-disk decomposition of massive CANDELS galaxies in the redshift range $0<z<2$, show that this is the case). Therefore, for a fair comparison of archaeological inferences with observed galaxies at different redshifts, instead of quiescent galaxies, galaxies with early-type morphologies should be selected. 
    
    \item The archaeological evolution of CLE galaxy progenitors is qualitatively similar for all masses but it differs in temporal scales, with less massive galaxies being roughly delayed versions of the massive ones. 
    
    \item According to our archaeological inferences, early-type galaxies observed at $z\sim 0.4-0.2$ (the smaller the mass, the lower $z$) should have, on average, more negative $M/L$ and colour gradients, as well as lower \rmass-to-\rlight\ ratios, than those observed at lower and higher redshifts. 
    { Regarding the latter, in Paper I we have shown that this result is in rough agreement with direct observations obtained at redshits from $z\sim 0$ \citep[e.g.,][]{Chan+2016} to $z\sim 1$ \citep[e.g.,][]{Suess+2019b}. As for the $M/L$ and color gradients, there are only a few studies, using relatively small samples. 
    For example, \citet{Ferreras+2005}, using the HST ACS imaging for the Southern GOODS field, showed that the $V-i$ color gradients for a set of early-type galaxies change from negative at $z\sim$ 1 to nearly flat or positive at $z\sim$ 0. 
  \citet{Menanteau+2004}, for a sample of 116 spheroidal galaxies in a broad range of masses at $z\lesssim 1$ found that a fraction of them have blue cores (positive gradients) and inhomogeneities on the color profiles, but they did not found a clear systematic trend of the gradient with $z$. 
  More observational studies are necessary.
}
\end{itemize}

The implications of our results partially challenge the evolutionary two-phases scenario for early-type galaxies \citep[e.g.,][]{Naab+2009,Oser+2010,Rodriguez-Gomez+2016}. According to this scenario and the observations that apparently support it \citep[e.g.,][and more references therein]{Daddi+2005,Trujillo+2006,Trujillo+2007,vanDokkum+2010,vanDokkum+2015,Cimatti+2012,Patel+2013,vanderWel+2014,Hill+2017,Mowla+2019}, these galaxies, after the early dissipative phase, suffered significant structural evolution due to dry minor mergers, increasing their sizes by factors $\sim3$ and being less and less compact since $z\sim 2$. The fossil record analysis presented here and in Paper I implies little \textit{intrinsic} structural evolution for most of the CLE galaxies. However, the $M/L$ ratio changes over time showing some size growth and concentration decrease in light.

It is important also to highlight that the fossil record method, at difference of look-back observational studies, reconstructs the 
evolution of { \textit{the same galaxy}}. As mentioned above, the progenitors of most of our analyzed CLE galaxies were forming stars actively in the not very distant past. In the look-back observational studies, those quiescent galaxies selected at high redshifts ($z\gtrsim 2$) to compare with those at low redshifts are probably  rare objects in the sense that formed in the highest density peaks and collapsed very early in the Universe, then being very compact. The more normal bulge-dominated galaxies at $z\gtrsim 1-2$ were probably still star-forming and less compact than those selected as quiescent at these high redshifts.\footnote{This argument is partially reinforced by the recent results by \citet[][]{Dimauro+2022} mentioned before \citep[see also][]{D'Eugenio+2021}. These authors conclude that growing a bulge is a required condition, but not sufficient to quench SF, which seems to imply that the bulge can grow before the quenching occurs, as they remark. } 
Therefore, identifying observed (rare) quiescent galaxies at high redshifts as progenitors of (more common) quiescent galaxies at lower redshifts may be strongly affected by a progenitor effect.  In addition to this, as mentioned in the Introduction, the look-back observational studies could lead to biased conclusions due to several observational and methodological effects (see the references therein). 

Recent studies based on selected samples from CANDELS fields at different redshifts, find that due to the $M/L$ and colour gradient changes with  $z$, the inferred mass-weighted size evolution of star-forming and quiescent galaxies is significantly less pronounced than that corresponding to the light-weighted size \citep[][]{Suess+2019b,Mosleh+2020,Hasheminia+2022,Miller+2022}, in qualitative agreement with our results. 
Also, a recent study of four massive quiescent galaxies around $z\sim2$ with available G141 deep slitless spectroscopic data and F160W photometric data from HST, has shown that their stellar age and metallicity radial profiles are consistent with those measured in local massive quiescent galaxies \citep[][]{Ditrani+2022}. 
In particular, the similarity of the negative metallicity gradients at these epochs with those retrieved in local galaxies suggest that the stellar populations of the passive galaxies are not redistributed spatially over the last $\sim10$ Gyr. According to the authors, this implies that the mechanisms that determine the main spatial distribution of the stellar population properties within passive galaxies occurred early in the dissipative phase.
If this lack of structural evolution applies for galaxies that were quiescent at those high redshifts, for galaxies that quench later, such a lack of evolution is even more likely.

\begin{figure}
\includegraphics[width=\columnwidth]{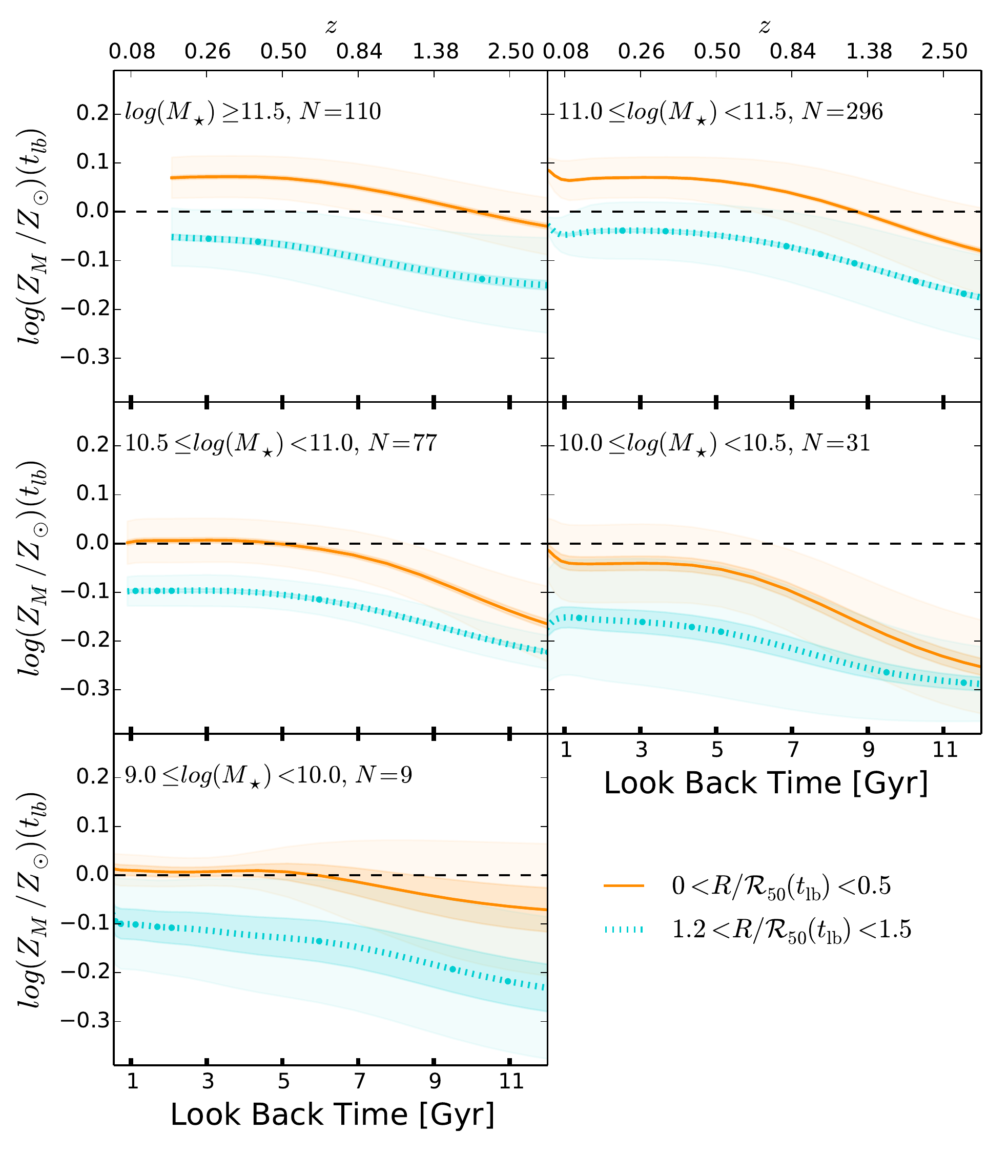}
    \caption{Evolution of the mass-weighted stellar metallicities within an inner and outer region of the galaxy archaeological progenitors. The median and 1st-3rd quartiles of $Z_{\star}^{\rm in}$ and $Z_{\star}^{\rm out}$, orange and cyan, respectively, are shown in the same five \ms($z_{\rm obs}$) bins from previous figures (in units of \msun). Both inner and outer metallicities remain almost constant since many Gyr ago. 
    }
    \label{fig:Z-evolution}
\end{figure}

More pieces of evidence in favor of passive evolution for quiescent massive galaxies since $z\sim 1$ come from the negligible evolution in the metallicity content found for them at $0.1<z<0.7$ \citep[][]{Choi+2014,Shetty-Cappellari2015} and from the age distribution found for a combination of massive quiescent galaxies at low and high redshifts \citep[][see also \citealp{Choi+2014}]{Mendel+2015}.
From the point of view of the fossil record reconstruction applied to the MaNGA survey, \citet[][]{Camps-Farina+2022} have shown that the light-weighted stellar metallicities of early-type galaxies remain roughly constant, on average, during the last $\sim 6-7$ Gyr for the most massive and $\sim 2.5-3$ Gyr for the least massive. We have calculated the mass-weighted stellar metallicity evolution in the inner ($<0.5\mathcal{R}_{50}$(\tlb)) and outer ($1.2-1.5\mathcal{R}_{50}$(\tlb)) regions for our CLE galaxies. 
The running medians and 25th-75th percentiles of $Z_{\star,M}^{\rm in}$ and  $Z_{\star,M}^{\rm out}$ in five \ms\ bins are shown in Fig. \ref{fig:Z-evolution}. On average, $Z_{\star,M}^{\rm in}$ remains roughly constant from \tlb$\sim 6-7$ Gyr for the more massive CLEs and from \tlb$\sim 4-5$ Gyr for the less massive ones. The outer regions follow more or less similar trends, but with lower values of $Z_{\star,M}$.  Therefore, the average metallicities of CLE progenitors do not change significantly, supporting this the claim that CLE galaxies evolved passively since the times mentioned above and quasi-passively at earlier times given that their metallicities increased only little since remote epochs. Furthermore, according to Fig. \ref{fig:Z-evolution}, the metallicity gradients should be negative since early epochs and do not change significantly over time, which is consistent with the lack of structural evolution for the progenitors of CLE galaxies after the dissipative phase, when the negative gradients were established \citep[e.g.,][]{Ditrani+2022}, at least within $\sim 1.5$ \rer.     
In a future paper, we will present a more detailed study of the evolution of metallicity and age gradients for the CLE galaxies, extending the analysis up to $\sim 2.5$ \rer\ for those galaxies from the Secondary sample.

\subsubsection{Implications for cosmological simulations}

The lack of structural evolution (at least since $z\approx 2-1$) 
of the progenitors of present-day quiescent elliptical galaxies that our fossil record analysis suggests is in tension with results from cosmological numerical simulations \citep{Pillepich2018MNRAS,Tacchella2019MNRAS,Pulsoni2021A&A,Cannarozzo+2023}, which endorse an evolutionary scenario in which the structure and outer growth of massive ETGs is affected by dry mergers.
Based on comparisons of the archaeological radial mass growth of zoom-in cosmological simulations of Milky Way-sized galaxies with fossil record inferences for MaNGA galaxies, \citet[][]{Avila-Reese+2018} pointed out to this potential tension, both for disc- and bulge-dominated galaxies. More recently, \citet[][]{Miller+2022} have compared the half-mass radius of their CANDELS galaxies over the redshift range $z=1-2$ with the radii reported in \citet[][]{Furlong+2017} for the EAGLE simulation and in \citet[][]{Genel+2018} for the TNG-100 simulation. The evolutionary trends in observations and simulations are qualitatively different for both star-forming and quiescent galaxies: in simulations the radii grow significantly,  while for observations the radii are not significantly different over this redshift range. In Paper I, we have shown that the recovered half-mass radii of our MaNGA CLE progenitors increase, on average, very weakly, in contrast to the strong growth seen for galaxies selected as quiescent at $z=0$ in the EAGLE simulation (Fig. 3 in \citealp[][]{Furlong+2017}). The size growth of galaxies in the cosmological simulations depends on the angular momentum acquirement and mass assembly of the dark matter haloes, and on the complex baryonic processes of galaxy evolution within these haloes, several of which are related to the assumed subgrid physics schemes and parameters \citep[see for a discussion][and more references therein]{Avila-Reese+2018}. Therefore, empirical information about the evolution of the radial stellar mass distribution of galaxies is crucial to constrain the subgrid physics for new generation of cosmological hydrodynamics simulations.

\section{Summary and Conclusions}
\label{sec:conclusions}

Using the fossil record method applied to 537 galaxies from the DR15 MaNGA IFS survey carefully classified as red and quiescent ellipticals, we have obtained 2D maps of stellar mass, light in different rest-frame bands, SFR, and stellar metallicities defined at different cosmic look-back times. 
In this way, we were able to reconstruct for these galaxies the \textit{individual} change over time (or redshift) of different radial gradients. Following the results from Paper I, our main goal here was to study the evolution of the mass and light radial distributions of the CLE galaxy progenitors characterized by the $M/L$ ratio gradient, $\nabla\Upsilon_\star$, as well as to explore the causes of its evolution. 

The main results of our study are summarized as follows:

\begin{itemize}
    \item The median $M/L$ gradients of the CLE progenitors start roughly flat at high look-back times (high redshfits), but then they become negative and steeper until an epoch, when this trend is reverted (Fig. \ref{fig:mass-light-grad-evol}). The more massive the CLE galaxies, the earlier $\nabla\Upsilon_\star$ starts to decrease and the earlier reaches the inversion; there is a trend of delayed evolution as lower is the mass. In the $r$ band, the median of $\nabla\Upsilon_{\star,r}$ reaches a minimum at \tlb$\sim 2.5$ Gyr ($z\sim 0.2$) for galaxies in the $9.0\le$log(\ms/\msun)$<10.0$ bin, and at \tlb$\sim$4 Gyr ($z\sim 0.35$) for those in the log(\ms/\msun)$\ge 11.5$ bin. 
    
    \item The evolutionary trends mentioned above for the $M/L$ gradients in different \ms\ bins are very similar to those of the \rmass-to-\rlight\ ratios reported in Paper I. On the other hand, the evolutionary trends of the $g-i$ colour gradients (Fig. \ref{fig:colour-grad-evol}) are qualitatively similar to those of the $M/L$ gradients, as expected.
    
    \item The median $\Upsilon_{\star,r}$ of CLE progenitors in different \ms\ bins increase rapidly from \tlb$\approx 8-5$ Gyr ago (the less massive the galaxy, the latter the epoch). The respective inner and outer $M/L$ ratios follow similar trends but $\Upsilon_{\star,r}^{\rm in}$ increases more rapidly than $\Upsilon_{\star,r}^{\rm out}$ until late epochs, when the rates become similar or even they invert for the less massive galaxies (Fig. \ref{fig:ML-evol-in-out}).
    
    \item The fossil record imprint in the observed CLE galaxies shows that, while their progenitors suffered strong inside-out quenching, any growth of the external stellar mass with respect to the internal is roughly the same since high redshifts. 
    The sSFR gradients inferred at different look-back times begin roughly flat at large \tlb, typical of star-forming galaxies, but then become positive, increasing the slope until they reach a maximum value at more recent epochs (Fig. \ref{fig:sSFRgradients}). We also calculate metrics for the change with time of the relative difference of the sSFRs and of the stellar mass surface  densities in an inner and an outer region. The more massive the galaxies, the earlier start to show less sSFR in the inner region than in the outer one. In contrast, the relative mass surface density differences remain roughly constant, evidencing little intrinsic structural changes (Fig. \ref{fig:inside-out-growth}). 
    
    \item CLE galaxies are dominated by old stellar populations ($>8$ Gyr). The fractions of these populations are, on average, only slightly higher in the inner regions than in the outer ones (Fig. \ref{fig:agefrac}). The fractions of stellar populations younger than 2 Gyr are very small, and they are on average $\sim 2\times$ higher in the outer regions than in the inner ones. 
    
    \item The scatters (1st-3rd quartile range) seen around the medians in \ms\ bins of the $M/L$ and sSFR gradients are large. We have shown that the individual random variations (variability) of the evolutionary trajectories contribute less to the scatter than the population dispersion. However, the individual variability is not negligible (Fig. \ref{fig:scatter}). 
\end{itemize}

We are aware that we are pushing the spatially-resolved spectroscopy data and its fossil record analysis to the limit. However, at a statistical level, thanks to the large number of objects and to the understanding of the effects of the observational/instrumental limitations and the uncertainties and degeneracies of the method (see \S\S\ \ref{sec:caveats}), the trends that we have found should be valid at a qualitatively level. 
These results and the different tests that we have carried out led us to conclude the following evolutionary scenario for the progenitors of present-day CLE galaxies.  After an early (dissipative) phase of intense SF and stellar mass growth, when the $M/L$ and sSFR gradients are roughly flat, a phase of SF quenching from inside out begins, earlier the   more massive the galaxies are, on average, steepening this the $M/L$ and colour gradients. 
Since we did not find major evidence of inside-out growth during this phase (or significant changes in stellar metallicity contents and gradients, Fig. \ref{fig:Z-evolution}),  most of the CLE progenitors had to have evolved in a quasi-passive regime, without significant structural changes (thought with important radial photometric changes). After the CLE progenitors quench globally and definitively, the light radial distribution tends to that of the stellar mass in such a way that the $M/L$ gradients attain a minimum and since then they increase slowly. Indeed, the times in which the minimum in the $M/L$ gradient trajectories are reached coincide approximately with the quenching times. As shown in Paper I, the 16th-84th percentiles of the quenching look-back time range from 4.8 to 2.6 Gyr for the more massive to less massive CLE galaxies, respectively. 

An implication of our results, as concluded in Paper I, is that while the half-light radii of CLE progenitors grows over time, their half-mass radii remain roughly constant. The above may explain partially the strong size growth claimed in studies based on photometric structural inferences of galaxies selected as quiescent at different redshifts (for references, see \S\S\ \ref{sec:implications}).  
Nonetheless, a major difference between these studies and our study is that { the fossil record method allows us to reconstruct the 
evolutionary trajectory of \textit{the same galaxy}}. 
The progenitors of { most of} our analyzed quiescent E galaxies were forming stars actively in the past but their radial stellar mass distributions did not change significantly since then, which implies (i) that the (inside-out) SF quenching happened after their structural-morphological transformation, and (ii) that galaxies selected as quiescent at higher redshifts in the look-back studies were not common objects (probably, were the most compact ones, formed in the highest density peaks at very early times). 
The latter suggests that selecting quiescent galaxies at different redshifts to infer the individual evolution of present-day galaxies of early-type morphologies may bias the result to show strong evolution from compact to extended, a trend that our archaeological inferences does not confirm { at a statistical level, though for a minor fraction of the analyzed galaxies this could be}. Our spatially-resolved archaeological inferences may offer also important constraints to the subgrid physics implemented in hydrodynamics simulations of galaxy evolution.

\section*{Acknowledgements}
We thank the referee for a careful and helpful review of our manuscript.  We thank Jimena A. Stephenson for providing us in electronic format with her data plotted in Fig. \ref{fig:colour-grad-evol}. This work
was supported by CONACyT ''Ciencia Basica'' 285721 grant and by UNAM-PAPIIT IN106823 grant.
HIM acknowledges a support grant from the Joint Committee ESO-Government of Chile (ORP 028/2020). IL acknowledges support from 'Proyecto DIUDA Programa Inserci\'on' N° 22414 of Universidad de Atacama'.

We acknowledge the SDSS-IV collaboration for making publicly available the data used in this paper. SDSS-IV is managed by the Astrophysical Research Consortium for the 
Participating Institutions of the SDSS Collaboration including the 
Brazilian Participation Group, the Carnegie Institution for Science, 
Carnegie Mellon University, the Chilean Participation Group, the French Participation Group, Harvard-Smithsonian Center for Astrophysics, 
Instituto de Astrof\'isica de Canarias, The Johns Hopkins University, 
Kavli Institute for the Physics and Mathematics of the Universe (IPMU)/University of Tokyo, 
Lawrence Berkeley National Laboratory, 
Leibniz Institut f\"ur Astrophysik Potsdam (AIP),  
Max-Planck-Institut f\"ur Astronomie (MPIA Heidelberg), 
Max-Planck-Institut f\"ur Astrophysik (MPA Garching), 
Max-Planck-Institut f\"ur Extraterrestrische Physik (MPE), 
National Astronomical Observatories of China, New Mexico State University, 
New York University, University of Notre Dame, 
Observat\'ario Nacional / MCTI, The Ohio State University, 
Pennsylvania State University, Shanghai Astronomical Observatory, 
United Kingdom Participation Group,
Universidad Nacional Aut\'onoma de M\'exico, University of Arizona, 
University of Colorado Boulder, University of Oxford, University of Portsmouth, 
University of Utah, University of Virginia, University of Washington, University of Wisconsin, 
Vanderbilt University, and Yale University.

This paper made use of the MaNGA-Pipe3D data products. We thank the IA-UNAM MaNGA team for creating this catalogue, and the CONACyT-180125 project for supporting them.

\section*{Data availability}

Most of the data underlying this article are available at the MaNGA-Pipe3D Valued Added Catalog at https://www.sdss.org/dr15/manga/manga-data/manga-pipe3d-value-added-catalog/. The datasets were derived from sources in the public domain using the SDSS-IV MaNGA public Data Release 15, at https://www.sdss.org/dr15/.




\bibliographystyle{mnras}
\bibliography{references-paperI.bib} 




\appendix


\begin{figure*}
	\includegraphics[width=1.8\columnwidth]{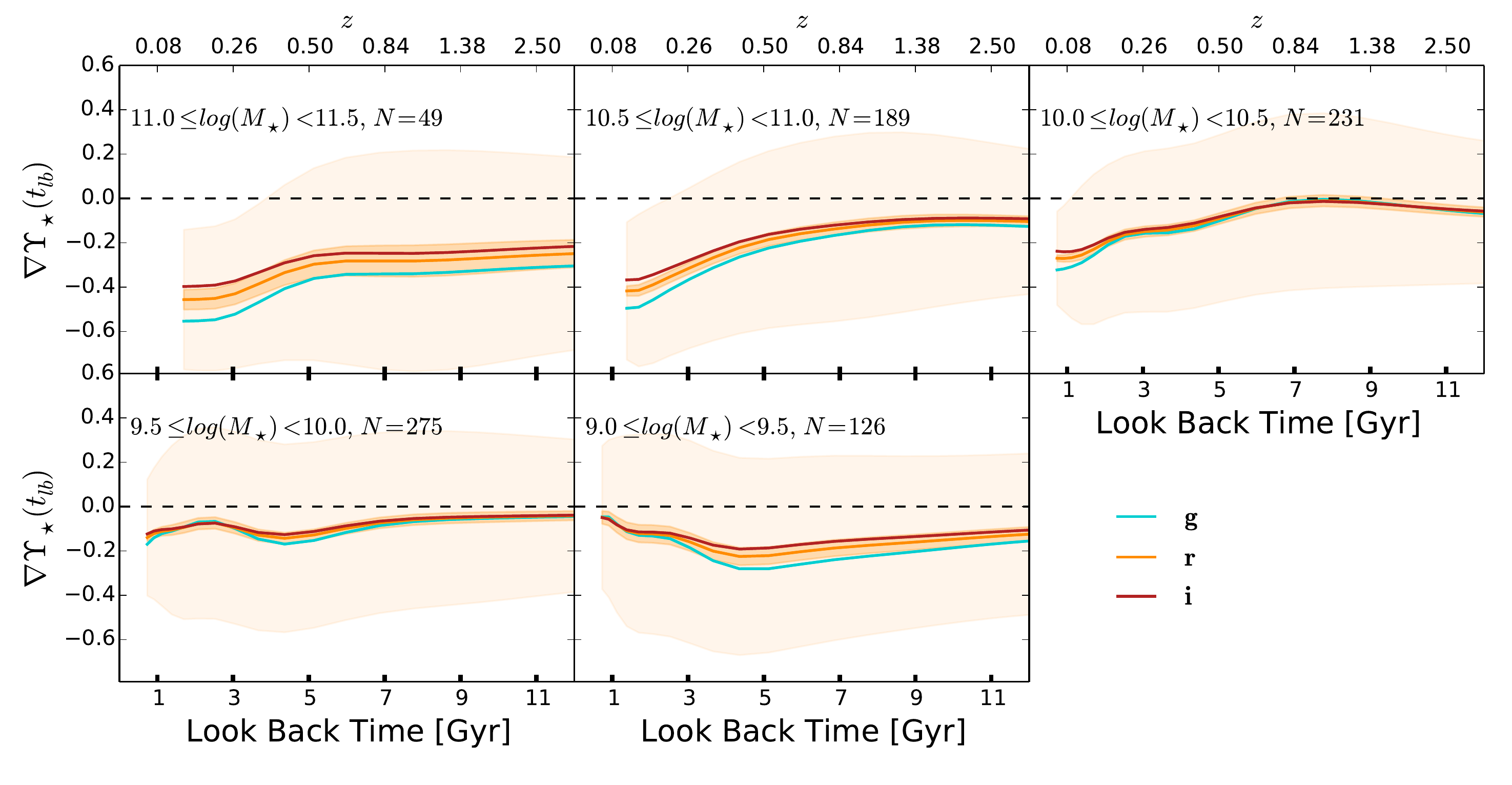}
    \caption{{ Evolution of the $M/L$  gradient of MaNGA spiral galaxies. Each panel shows the running medians of the gradients for the $g,$ $r,$ and $i$ bands in the \ms($z_{\rm obs})$\ bins shown inside the panels (in units of \msun). The light shaded regions correspond to the associated 25th-75th percentiles for the $r$ band, while the dark shaded regions are estimates of the standard error of the median. }
    }
    \label{fig:spirals}
\end{figure*}


\section{The $M/L$ gradient evolution of spiral galaxies}
\label{sec:appendix-spirals}
{ 
The inferred $M/L$ gradient evolution for the CLE galaxies systematically attains a minimum at relatively late epochs and then tends to slightly flattens (see Fig. \ref{fig:mass-light-grad-evol}), something that we interpreted in \S\S\ \ref{sec:inside-out} as a signature of global and definitive SF quenching. It could be that this characteristic blip in the $M/L$ gradient evolution is associated to an artefact of our fossil record methodology.  If this would be the case instead of a true evolutionary effect intrinsic to the CLE galaxies (as we interpret in \S\S\ \ref{sec:inside-out}), a similar characteristic blip would appear for other galaxy types. 
Thus, we have also calculated the $M/L$ gradient evolution for MaNGA DR15 spiral star-forming galaxies. Galaxies later or equal than Sb according to our visual morphological classification, obeying the criteria of star-forming, and less inclined than 60 degrees were selected for the analysis. Figure \ref{fig:spirals} shows the median archaeological evolution of the $M/L$ gradient  in  five \ms\ bins, similar to Fig. \ref{fig:mass-light-grad-evol}. There are only a few galaxies with log(\ms/\msun)$>11.5$, so we do not show this bin. On the other hand, most of galaxies have masses below $10^{10}$ \msun, so we divided them into two mass bins.  
For galaxies more massive than $\sim 10^{10}$ \msun, the results show that the $M/L$ gradients continue decreasing up to very recent epochs for the massive ones or to the corresponding observation redshift for the less massive ones. This is the expected evolutionary trend as discussed below. 

On the one hand, spiral galaxies 
are still star forming (though for the more massive a significant decline in SFR at all radii may have started recently), so their $M/L$ gradients at the observation redshift are expected to be negative and steep. 
Furthermore, it is well known that spiral galaxies show clear trends of inside-out mass growth in such a way that their $M/L$ gradients should decrease continuously over time. As for the low-mass spirals, \ms$<10^{10}$ \msun, they have in general a delayed SF history, so they can be in the initial phases of decreasing their $M/L$ gradients. On the other hand, the lower the mass, the more susceptible the galaxies are to SF feedback effects (e.g., fountain processes), which could affect the radial evolution of these galaxies and explain the big scatter in the $M/L$ gradients of low-mass spirals.  
It is out of the scope of this paper to study the evolution of radial gradients of spiral galaxies, but we presented these results to show that the late $M/L$ gradient blip found for CLE galaxies is intrinsic and not a systematic artifact of our methodology, in which case it would also be seen for other galaxies.

}

\section{Inner and outer fractions of stellar populations by their ages}
\label{sec:appendix}

We compute here the stellar mass fractions,$f_M$,  contained in stellar populations of three age ranges, both for an inner ($<0.5$ \rer) and outer (between 1.2 and 1.5 \rer) region of our CLE galaxies. The age ranges are Age$>8$ Gyr (old), $2<$Age/Gyr$\le 8$ (intermediate), and Age$\le 2$ Gyr (young), and the fractions are with respect to the stellar mass within each region. 

\begin{figure}
\includegraphics[width=0.9\columnwidth]{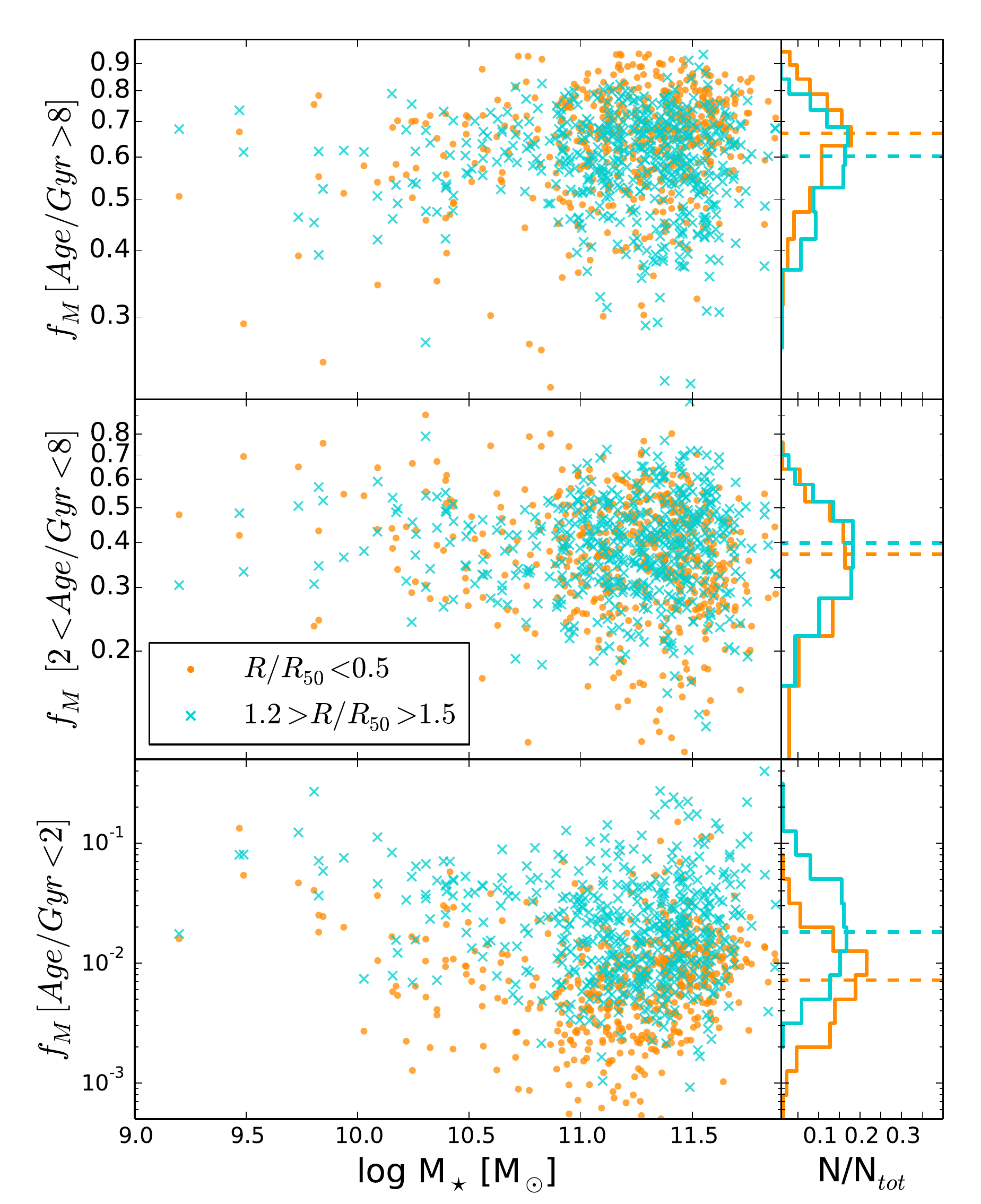}
    \caption{Mass fractions of stellar populations older than 8 Gyr (upper panel), between 2 and 8 Gyrs (midle panel), and younger than 2 Gyr (lower panel), in an inner region ($<$0.5 \rer; orange dots) and in an outer region (1.2--1.5 \rer; cyan crosses) as a function of \ms($z_{\rm obs}$). The fractions are with respect to the total masses in these regions. Note that the y-axis ranges change in each panel. The right panels show the respective normalized histograms of both fractions. The horizontal dashed lines are the median of the distributions. 
    }
    \label{fig:agefrac}
\end{figure}

Figure \ref{fig:agefrac} shows the different mass fractions for the inner and outer regions, orange dots and cyan crosses, respectively, as a function of \ms. The distributions of the different fractions are shown in the side panels, with the dashed lines showing the medians corresponding to the inner and outer regions. 
The distribution of the mass fraction of old stars ($>$8 Gyr) is only slightly shifted to higher values in the inner regions with respect to the the outer one; the medians and 16th-84th percentiles are $0.67^{+0.12}_{-0.15}$ versus $0.60^{+0.11}_{-0.14}$, respectively. At the individual level, for $\approx 66\%$ ($\approx 34\%$) of the CLEs, the fraction of old stars in the outer region is smaller (larger) than in the inner one. In the opposite side, the mass fractions of young stars ($<$2 Gyr), are very low in general, being higher in the outer regions than in the inner ones, $0.02^{+0.02}_{-0.01}$ versus $0.01^{+0.008}_{-0.004}$.
The fractions of intermediate age stars (2-8 Gyr), are in general smaller than those of old stars and are roughly the same in the inner and outer regions,  $0.38^{+0.15}_{-0.11}$ versus $0.40^{+0.13}_{-0.11}$.  
In summary, CLE galaxies of all masses are dominated by old stellar populations with only small differences in fractions between inner ($<$0.5 \rer) and outer (1.2--1.5 \rer) regions.

\bsp	
\label{lastpage}

\end{document}